%% file: arxiv-v2.tex
\newcommand\withproofs

\newcommand\arxiv

\ifdefined \arxiv
\providecommand\usefullpage
\fi

\documentclass{llncs}

 \usepackage{times}

\usepackage{wrapfig}

\ifdefined \techreport
\providecommand\usefullpage
\fi

\ifdefined \usefullpage
\usepackage{fullpage}
\fi

\usepackage{subcaption}

\usepackage{enumitem}

\usepackage{color}
\usepackage[colorlinks,naturalnames,citecolor=blue,urlcolor=blue]{hyperref}
\newcommand{\myhref}[1]{\href{#1}{\url{#1}}}

\newcommand{\trainset}[1]{R_{\sf{train}}}
\newcommand{\testset}[1]{R_{\sf{test}}}

\usepackage{multirow}

\usepackage{tikz}
\usetikzlibrary{calc,intersections,through,scopes,arrows,decorations,backgrounds,positioning,fit,petri,automata,shapes}

\usepackage{float}

\usepackage{listings}

\DeclareFixedFont{\ttb}{T1}{txtt}{bx}{n}{8} 
\DeclareFixedFont{\ttm}{T1}{txtt}{m}{n}{8}  

\begin{document}

\ifdefined \arxiv
\interfootnotelinepenalty=10000
\fi

\title{
Learning Moore Machines from Input-Output Traces\thanks{
This work was partially supported by the Academy of Finland and the U.S. National Science Foundation (awards \#1329759 and \#1139138).
\ifdefined \arxiv
\else
An extended version of this paper is available as~\cite{LearningMooreArxiv2016}.
\fi
}
}

\author{Georgios Giantamidis\inst{1} \and Stavros Tripakis\inst{1,2}}
\institute{Aalto University, Finland
\and
University of California, Berkeley, USA
}

\maketitle

\begin{abstract}
The problem of learning automata from example traces (but no equivalence or membership queries) is fundamental in automata learning theory and practice.
In this paper we study this problem for finite state machines with
inputs and outputs, and in particular for Moore machines. 
We develop three
algorithms for solving this problem: (1) the PTAP algorithm, which
transforms a set of input-output traces into an incomplete 
Moore machine
and then completes the machine with self-loops; (2) the PRPNI algorithm,
which uses the well-known RPNI algorithm for automata learning to learn
a product of automata encoding a Moore machine; and (3) 
the MooreMI algorithm, which directly learns a Moore machine using PTAP extended with state merging.
We prove that MooreMI 
has the fundamental {\em identification in the limit} property.
We also compare the algorithms experimentally in terms of the size of the learned machine and several notions of accuracy, introduced in this paper. Finally, we compare with OSTIA, an algorithm that learns a more general class of transducers, and find that OSTIA generally does not learn a Moore machine, even when fed with a {\em characteristic sample}.
\end{abstract}

\section{Introduction}

An abundance of data from the internet and from other sources (e.g.,
sensors) is revolutionizing many sectors of science, technology, 
and ultimately our society. At the heart of this revolution lies 
{\em machine learning}, a broad spectrum of techniques to derive information
from data. Traditionally, objects studied by machine learning include
classifiers, decision trees, and neural networks, with applications to 
fields as diverse as artificial intelligence,
marketing, finance, or medicine~\cite{Mitchell1997}.

In the context of system design, an important problem, with numerous
applications, is automatically generating models from data. There are
many variants of this problem, depending on what types of models and
data are considered, as well as other assumptions or restrictions.
Examples include, but are by no means limited to, the classic field of
system identification~\cite{Ljung1999}, as well as more recent works
on synthesizing programs, controllers, or other artifacts from examples~\cite{SolarLezama13,Gulwani11,seshia-dac12,ray2014lang,ScenariosHVC2014}.

In this paper we consider a basic problem, that of learning a Moore
machine from a set of input-output traces. A Moore machine is a type of
finite-state machine (FSM) with inputs and outputs, where
the output always depends on the current state, but not on the current
input~\cite{Kohavi78}.
Moore machines are typically {\em deterministic} and {\em complete},
meaning that for given state and input, the next state is always defined
and is unique; and for given state, the output is also always uniquely
defined. Such machines are useful in many applications, for instance,
for representing digital circuits or controllers.
In this paper we are interested in learning
deterministic and complete Moore machines.

We want to learn a Moore machine from a given set of 
{\em input-output traces}. One such trace is a sequence of inputs, 
$\rho_{in}$, and the corresponding sequence of outputs, $\rho_{out}$,
that the machine must produce when fed with $\rho_{in}$.
As in standard machine learning methods, we call the set of traces given
to the learning algorithm the {\em training} set. Obviously, 
we would like the learned machine $M$ to
be {\em consistent} w.r.t. the training set $R$, meaning that for every
pair $(\rho_{in},\rho_{out})\in R$, $M$ must output $\rho_{out}$ when
fed with $\rho_{in}$. But in addition to consistency, we would like $M$ to 
behave well w.r.t. several {\em performance} criteria, including complexity
of the learning algorithm, size of the learned machine $M$
(its number of states), and {\em accuracy} of $M$, which captures how well 
$M$ performs on a {\em testing} set of traces, different from the training set.

Even though this is a basic problem, it appears not to have
received much attention in the literature. 
In fact, to the best of our knowledge, this is the first paper which 
formalizes and studies this problem.
This is despite a large body of research on
{\em grammatical inference}~\cite{delaHiguera2010} which has studied similar,
but not exactly the same problems, such as learning deterministic finite
automata (DFA), which are special cases of Moore machines with a binary
output, or subsequential transducers, which are more general than Moore machines.

Our contributions are the following:
\begin{enumerate}[leftmargin=*]
\item
We define formally the LMoMIO problem (learning Moore machines from
input-output traces).
Apart from the correctness criterion of {\em consistency} (that the learned
machine be consistent with the given traces) we also introduce several
{\em performance} criteria including size and accuracy of the learned machine,
and computational complexity of the learning algorithm.

\item 
 We adapt the notion of {\em characteristic sample}, which is known for
DFA~\cite{delaHiguera2010}, to the case of Moore machines. 
Intuitively, a characteristic sample of a machine $M$ is a set of traces
which contains enough information to ``reconstruct'' $M$.
The {\em characteristic sample requirement} (CSR) states that, when given as input
a characteristic sample, the learning algorithm must produce a machine 
equivalent to the one that produced the sample.
CSR is important, as it ensures {\em identification in the limit}: this is
a key concept in automata learning theory which ensures that the learning
algorithm will eventually learn the right machine when provided with a
sufficiently large set of examples~\cite{DBLP:journals/iandc/Gold67}.

\item

We develop three algorithms to solve the LMoMIO problem, and
analyze them in terms of computational complexity and other properties. We show that although all three algorithms guarantee consistency,
only the most advanced among them, called {\em MooreMI},
satisfies the characteristic sample requirement.
We also show that MooreMI achieves identification in the limit.

\item
We report on a prototype implementation of all three algorithms and 
experimental results. The experiments show that MooreMI outperforms the other two algorithms not only in theory, but also in practice.

\item
We show that the well-known 
transducer-learning algorithm 
OSTIA~\cite{DBLP:journals/pami/OncinaGV93} cannot generally learn a Moore
machine, even in the case where the training set is a characteristic sample
of a Moore machine.
This 
implies that an algorithm to learn a more general machine
(e.g., a transducer)
is not necessarily good at learning a more special machine, and therefore 
further justifies 
the study of specialized learning algorithms for Moore machines.
\end{enumerate}

\section{Related Work}
\label{sec_related}

There is a large body of research on learning automata and state machines,
which
can be divided into two broad categories: learning with 
(examples and) queries ({\em active} learning), and learning only from examples
({\em passive} learning).
A seminal work in the first category is 
Angluin's work on learning DFAs with membership and equivalence queries~\cite{Angluin1987}. 
This work has been subsequently extended to other types of machines,
such as Mealy machines~\cite{DBLP:conf/fm/ShahbazG09}, symbolic / extended Mealy machines~\cite{DBLP:conf/sfm/Jonsson11,DBLP:conf/sefm/CasselHJS14}, I/O automata~\cite{Aarts2010}, register automata~\cite{DBLP:conf/vmcai/HowarSJC12,DBLP:conf/ictac/AartsFKV15}, or hybrid automata~\cite{MedhatRBF_EMSOFT15}.
These works are not directly applicable to the problem studied in this paper,
as we explicitly forbid both membership and equivalence queries.
In practice, performing queries (especially complete
equivalence queries) is often infeasible.

In the domain of passive learning, a seminal work is Gold's study of learning DFAs from sets of positive and negative examples~\cite{DBLP:journals/iandc/Gold67,DBLP:journals/iandc/Gold78}.
In this line of work we must distinguish algorithms that solve
the {\em exact identification} problem, which is to find a {\em smallest}
(in terms of number of states) automaton consistent with the given examples,
from those that learn not necessarily a smallest automaton\footnote{
The term {\em smallest} automaton is used in the exact identification problem, 
instead of the more well-known term {\em minimal} automaton.
Among equivalent machines, one with the fewest states is called {\em minimal}.
Among machines which are all consistent with a set of traces but not
necessarily equivalent, one with the fewest states is called {\em smallest}.
}
(let us call them {\em heuristic} approaches).
 Gold showed that exact identification is NP-hard for DFAs~\cite{DBLP:journals/iandc/Gold78}.
Several works solve the exact identification problem by
reducing it into boolean 
satisfiability~\cite{Heule:2013:SMS:2506969.2506989,ulyantsev/bfssym}.

Heuristic approaches are dominated by state-merging algorithms like Gold's algorithm for DFAs \cite{DBLP:journals/iandc/Gold78}, RPNI \cite{Oncina92identifyingregular} (also for DFAs),
for which an incremental version also exists \cite{DBLP:conf/icgi/Dupont96}, and derivatives, like EDSM \cite{DBLP:conf/icgi/LangPP98} (which also learns DFAs, but unlike RPNI does not guarantee identification in the limit) and OSTIA \cite{DBLP:journals/pami/OncinaGV93} (which learns subsequential transducers).
This line of work also includes gravitational search algorithms~\cite{gravity.moore},
genetic algorithms~\cite{Aleksandrov2013}, ant colony optimization~\cite{Buzhinsky2014}, 
rewriting~\cite{DBLP:conf/icgi/Meinke10},
as well as state splitting algorithms~\cite{DBLP:journals/tc/Veelenturf78}.
\cite{gravity.moore} learns Moore machines, but unlike our work does not guarantee identification in the limit.~\cite{DBLP:journals/tc/Veelenturf78,DBLP:conf/icgi/Meinke10,Aleksandrov2013,Buzhinsky2014} all learn Mealy machines.

All algorithms developed in this paper belong in the heuristic category
in the sense that we do not attempt to find a {\em smallest} machine. However,
we would still like to learn a {\em small} machine. Thus, size is an important
{\em performance} criterion, as explained in Section~\ref{sec_problem}.
Like RPNI and other algorithms, MooreMI is also a state-merging algorithm.

\cite{DBLP:journals/scjapan/TakahashiFK03} is close to our work, but the algorithm described there does not always yield a deterministic Moore machine, while our algorithms do.
This is important because we want to learn systems like digital circuits, embedded controllers (e.g. modeled in Simulink), etc., and such systems are typically deterministic.
The k-tails algorithm for finite state machine inference~\cite{Biermann:1972:SFM:1638603.1638997} may also result in non-deterministic machines.
Moreover, this algorithm does not generally yield smallest machines, 
since the initial partition of the input words into equivalence classes (which then become the states of the learned machine) can be overly conservative
\ifdefined \arxiv
.\footnote{We have implemented the k-tails algorithm and applied it on the characteristic sample for the Moore machine in Figure~\ref{figwhiteboardex1}, described in Section~\ref{sec_cs}. Using $k = 0$, we get a non-deterministic machine of 3 states. Using any $k > 0$, we get a deterministic machine of 8 states. This excessive number of states is due to the way the k-tails equivalence relation is defined. In particular, in order for two input words to be considered equivalent, they must have successors in the training set with the same letters. This implies that a word with no successors in the training set can never be equivalent with a word with some successors, even if both words represent the same state in the target machine.}
\else
(see~\cite{LearningMooreArxiv2016} for details).
\fi

The work in~\cite{6742898} deals with learning finite state machine abstractions of non-linear analog circuits. The algorithm 
described in~\cite{6742898} is very different from ours, and uses
the circuit's number of inputs to determine a subset of the states in the 
learned abstraction. Also, identification in the limit is not considered
in~\cite{6742898}. 

Learning from ``inexperienced teachers'', i.e. by using either (1) only equivalence queries or (2) equivalence plus membership queries that may be answered inconclusively, has been studied
\ifdefined \arxiv
in~\cite{DBLP:conf/icgi/GrinchteinL06,DBLP:conf/isola/LeuckerN12}.
\else
in~\cite{DBLP:conf/icgi/GrinchteinL06}.
\fi

\ifdefined \techreport

\fi

Related but different from our work are approaches which synthesize
state machines from {\em scenarios and requirements}. Scenarios can be provided in various forms, e.g. message sequence charts \cite{ScenariosHVC2014}, event sequence charts \cite{Heitmeyer2015}, or simply, input-output examples \cite{DBLP:journals/corr/UlyantsevBS16}. Requirements can be temporal logic formulas
as in~\cite{ScenariosHVC2014,DBLP:journals/corr/UlyantsevBS16},
or other types of constraints such as the {\em scenario constraints} used
in~\cite{Heitmeyer2015}.
In this paper we have examples, but no requirements.

Also related but different from ours is work in the areas of {\em invariant generation} and {\em specification mining}, which extract properties of a program
or system model, such as invariants \cite{Gulwani:2008:PAC:1375581.1375616,inv.gen.cav2003,inv.gen.cav2009}, temporal logic formulas \cite{seshia/donze:req/mining,caroline:ltl/spec/mining} or non-deterministic finite automata~\cite{Ammons:2002:MS:503272.503275}.

FSM learning is related to FSM testing~\cite{lee.yannakakis.testing.fsms}. In particular, notions similar to the {\em nucleus} of an FSM and to {\em distinguishing suffixes} of states, which are used to define characteristic samples, 
are also used in~\cite{Chow:1978:TSD:1313335.1313730,Dorofeeva:2010:FCT:1864811.1864946}. 
The connection between conformance testing and regular inference is made
explicit in~\cite{DBLP:conf/fase/BergGJLRS05}
and~\cite{lee.yannakakis.testing.fsms} describes how an active learning algorithm can be used for fault detection.

Reviewers of an earlier version of this paper pointed out the similarity of
Moore and Mealy machines: a Moore machine is a special case of a Mealy machine
where the output depends only on the state but not on the input; and a Mealy
machine can be transformed into a Moore machine by delaying the output by
one step.
This similarity naturally raises the question to what extent methods to
learn Mealy machines can be used to learn Moore machines (and vice versa).
Answering this question is beyond the scope of the current
paper. However,
note that an algorithm that learns a Mealy machine cannot be used as a
{\em black box} to learn Moore machines, for two reasons:
first, the input-output
traces for a Moore machine are not directly compatible with Mealy machines,
and therefore need to be transformed somehow; second, the learned Mealy
machine must also be transformed into a Moore machine. The exact form of
such transformations and their correctness remain to be demonstrated.
Such transformations may also incur performance penalties 
which make a learning method especially designed for Moore machines more
attractive in practice.

\section{Preliminaries}

\subsection{Finite state machines and automata}
 
A {\em finite state machine} (FSM) is a tuple $M$ of the form
$M=(I,O,Q,q_0,\delta,\lambda)$, where:
\ifdefined \arxiv
\begin{itemize}
\item $I$ is a finite set of {\em input symbols}.
\item $O$ is a finite set of {\em output symbols}.
\item $Q$ is a finite set of {\em states}.
\item $q_0\in Q$ is the {\em initial state}.
\item $\delta:Q\times I\to Q$ is the {\em transition function}.
\item $\lambda$ is the {\em output function}, which can be of two types:
\begin{itemize}
\item $\lambda:Q\to O$, in which case the FSM is a {\em Moore machine}.
\item $\lambda:Q\times I\to O$, in which case the FSM is a {\em Mealy machine}.
\end{itemize}
\end{itemize}
\else
$I$ is a finite set of {\em input symbols};
$O$ is a finite set of {\em output symbols};
$Q$ is a finite set of {\em states};
$q_0\in Q$ is the {\em initial state};
$\delta:Q\times I\to Q$ is the {\em transition function};
and
$\lambda$ is the {\em output function}, which can be of two types:
$\lambda:Q\to O$, in which case the FSM is a {\em Moore machine}, or
$\lambda:Q\times I\to O$, in which case the FSM is a {\em Mealy machine}.
\fi

\begin{figure}[t]
  \begin{subfigure}[t]{.49\columnwidth}
    \centering
    \ifdefined\pdfonlyfigs
	\includegraphics[]{figures/moore_o.pdf}
	\else
	\resizebox{0.55\textwidth}{!}{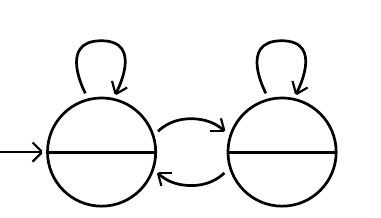}
    \fi
    \caption{Moore machine $M_1$ on input-output sets
	$I = \{x_1, x_2\}$ and $O=\{y_1,y_2\}$.}
    \label{figmooreexample}
  \end{subfigure}
  \quad
  \begin{subfigure}[t]{.49\columnwidth}
    \centering
    \ifdefined\pdfonlyfigs
	\includegraphics[]{figures/mealy_o.pdf}
	\else
	\resizebox{0.55\textwidth}{!}{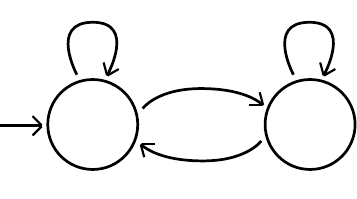}
    \fi
    \caption{Mealy machine $M_2$ on input-output sets
	$I = \{x_1, x_2\}$ and $O=\{y_1,y_2\}$.}
    \label{figmealyexample}
  \end{subfigure}
  \caption{Examples of finite state machines.}
  \label{figFSMexamples}
\end{figure}

If both $\delta$ and $\lambda$ are total functions, we say that the FSM is {\em complete}. 
If any of $\delta$ and $\lambda$ is a partial function, we say that the FSM is {\em incomplete}.
Examples of a Moore and a Mealy machine are given in Figure~\ref{figFSMexamples}.
Both FSMs are complete.

We also define $\delta^*:Q\times I^*\to Q$ as follows ($X^*$ denotes the set of all finite sequences over some set $X$; $\epsilon\in X^*$ denotes the empty sequence over $X$; $w\cdot w'$ denotes the concatenation of two sequences $w,w'\in X^*$):
for $q\in Q$, $w\in I^*$, and $a\in I$:
\ifdefined \arxiv
\begin{itemize}
\item $\delta^*(q, \epsilon) = q$.
\item $\delta^*(q, w\cdot a) = \delta(\delta^*(q, w), a)$.
\end{itemize}
\else
$\delta^*(q, \epsilon) = q$ and 
$\delta^*(q, w\cdot a) = \delta(\delta^*(q, w), a)$.
\fi
We also define $\lambda^*:Q\times I^*\to O^*$.
The rest of this paper focuses on Moore machines, thus we
define $\lambda^*$ only in the case where $M$ is a Moore machine
(the adaptation to a Mealy machine is straightforward):
\ifdefined \arxiv
\begin{itemize}
\item $\lambda^*(q, \epsilon) = \lambda(q)$
\item $\lambda^*(q, w\cdot a) = \lambda^*(q, w) \cdot \lambda(\delta^*(q, w \cdot a)) $ 
\end{itemize}
\else
$\lambda^*(q, \epsilon) = \lambda(q)$ and
$\lambda^*(q, w\cdot a) = \lambda^*(q, w) \cdot \lambda(\delta^*(q, w \cdot a))$.
\fi

Two Moore machines $M_1,M_2$, with $M_i=(I_i,O_i,Q_i,q_{0\_i},\delta_i,\lambda_i)$, are said to be {\em equivalent} iff $ I_1 = I_2 $, $ O_1 = O_2 $, and
$ \forall w \in I_1^* : \lambda_1^*(q_{0\_1}, w) =  \lambda_2^*(q_{0\_2}, w)$.

A Moore machine $M=(I,O,Q,q_0,\delta,\lambda)$ is {\em minimal} if for
any other Moore machine $M'=(I',O',Q',q_0',\delta',\lambda')$ such that $M$ and $M'$ are equivalent, we have $|Q| \leq |Q'|$, where $|X|$ denotes the size of a set $X$.

\ifdefined \arxiv
Notice that in the case two Moore machines are minimal, testing equivalence is reduced to a graph isomorphism test.
\fi

A {\em deterministic finite automaton} (DFA) is a tuple
$A=(\Sigma,Q,q_0,\delta,F)$,
\ifdefined \arxiv
where:
\begin{itemize}
\item $\Sigma$ (the {\em alphabet}) is a finite set of {\em letters}.
\item $Q$ is a finite set of {\em states}.
\item $q_0\in S$ is the {\em initial state}.
\item $\delta:Q\times \Sigma\to Q$ is the {\em transition function}.
\item $F\subseteq Q$ is the set of {\em accepting states}.
\end{itemize}
\else
where:
$\Sigma$ (the {\em alphabet}) is a finite set of {\em letters};
$Q$ is a finite set of {\em states};
$q_0\in S$ is the {\em initial state};
$\delta:Q\times \Sigma\to Q$ is the {\em transition function};
$F\subseteq Q$ is the set of {\em accepting states}.
\fi

A DFA can be seen as a special case of a Moore machine, where the set of
input symbols $I$ is $\Sigma$, and the set of output symbols is binary,
say $O=\{0,1\}$, with $1$ and $0$ corresponding to accepting and non-accepting
states, respectively.
The concepts of {\em complete} and {\em incomplete} DFAs, as well as the definition of $\delta^*$, are similar to the corresponding ones for FSMs.
Elements of $\Sigma^*$ are usually called {\em words}. 
A DFA $A=(\Sigma,Q,q_0,\delta,F)$ is said to accept a word $w$ if $\delta^*(q_0, w) \in F$.

\begin{figure}[t]
  \begin{subfigure}[t]{.49\columnwidth}
    \centering
    \ifdefined\pdfonlyfigs
	\includegraphics[]{figures/dfa_o.pdf}
	\else
	\resizebox{0.47\textwidth}{!}{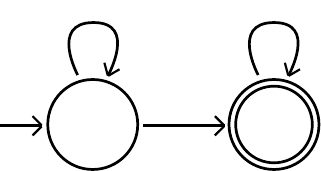}
    \fi
    \caption{DFA $A_1$ on $\Sigma = \{a, b\}$.}
    \label{figdfaexample}
  \end{subfigure}%
  \begin{subfigure}[t]{.49\columnwidth}
    \centering
    \ifdefined\pdfonlyfigs
	\includegraphics[]{figures/nfa_o.pdf}
	\else
	\resizebox{0.47\textwidth}{!}{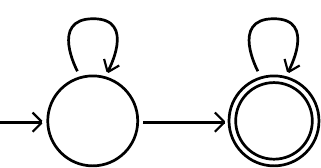}
    \fi
    \caption{NFA $A_2$ on $\Sigma = \{a, b\}$.}
    \label{fignfaexample}
  \end{subfigure}
  \caption{Examples of finite state automata.}
  \label{figautomataexamples}
\end{figure}

A {\em non-deterministic finite automaton} (NFA) is a tuple
$A=(\Sigma,Q,Q_0,\Delta,F)$, where
$\Sigma$, $Q$, and $F$ are as in a DFA, and:
\ifdefined \arxiv
\begin{itemize}
\item $Q_0\subseteq Q$ is the {\em set of initial states}.
\item $\Delta\subseteq Q\times \Sigma\times Q$ is the {\em transition relation}.
\end{itemize}
\else
$Q_0\subseteq Q$ is the {\em set of initial states};
$\Delta\subseteq Q\times \Sigma\times Q$ is the {\em transition relation}.
\fi
Examples of a DFA and an NFA are given in Figure~\ref{figautomataexamples}.
Accepting states are drawn with double circles.

Given two NFAs, $A_1=(\Sigma,Q_1,Q_0^1,\Delta_1,F_1)$ and
$A_2=(\Sigma,Q_2,Q_0^2,\Delta_2,F_2)$, their {\em synchronous product}
is the NFA $A=(\Sigma,Q_1\times Q_2, Q_0^1\times Q_0^2, \Delta, F_1\times F_2)$,
where $((q_1,q_2),a,(q_1',q_2'))\in\Delta$ iff 
$(q_1,a,q_1')\in\Delta_1$ and $(q_2,a,q_2')\in\Delta_2$. 
The synchronous product of automata is used
in several algorithms presented in the sequel.

\subsection{Input-output traces and examples}
 
Given sets of input and output symbols $I$ and $O$, respectively,
a
{\em Moore $(I,O)$-trace}
 is a pair of finite sequences
$(x_1 x_2 \cdots x_n,\; y_0 y_1 \cdots y_n)$, for some natural number $n\ge 0$,
such that $x_i\in I$ and $y_i\in O$ for all $i\le n$.
That is, a Moore $(I,O)$-trace is a pair of a input sequence and an output
sequence, such that the output sequence has length one more than the input
sequence.
Note that $n$ may be $0$, in which case the input sequence is empty
(i.e., has length $0$), and the output sequence contains just one output symbol.

Given a Moore $(I,O)$-trace $\rho=(x_1 x_2 \cdots x_n,\; y_0 y_1 \cdots y_n)$,
and a Moore machine $M=(I,O,Q,q_0,\delta,\lambda)$, we say that
{\em $\rho$ is consistent with $M$} if $y_0 = \lambda(q_0)$ and for all
$i = 1,...,n$, $y_i = \lambda(q_i)$, where $q_i = \delta(q_{i-1},x_i)$.

Similarly to the concept of a Moore $(I,O)$-trace we define a {\em Moore $(I,O)$-example} as a pair of a finite input symbol sequence and an output symbol: $(x_1 x_2 \cdots x_n,\; y)$, where $x_i\in I$, for $i=1,...,n$, and $y\in O$. We say that a Moore machine $M=(I,O,Q,q_0,\delta,\lambda)$ is consistent with a Moore $(I,O)$-example $\rho=(x_1 x_2 \cdots x_n,\; y)$ if $\lambda(\delta^*(q_0, x_1 x_2 \cdots x_n))=y$.

Since a DFA can be seen as the special case of a Moore machine with a binary output alphabet, the concept of a Moore $(I,O)$-example is naturally carried over to DFAs, in the form of {\em positive and negative examples}. 
Specifically, a finite
word $w$ is a {\em positive} example for a DFA if it is accepted
by the DFA, and a {\em negative} example if it is rejected.
Viewing a DFA as a Moore machine with binary output, a positive example $w$
corresponds to the Moore example $(w,1)$, while a negative example corresponds
to the Moore example $(w,0)$.

\subsection{Prefix tree acceptors and prefix tree acceptor products}
\label{sec_ptas}

\ifdefined \arxiv
\begin{wrapfigure}{r}{.35\textwidth}
	\vspace{-45pt}
    \centering
    \ifdefined\pdfonlyfigs
	\includegraphics[]{figures/pta_o.pdf}
	\else
	\resizebox{0.31\columnwidth}{!}{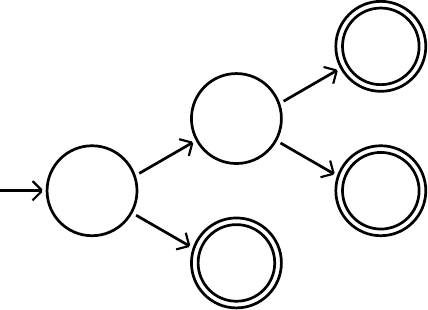}
    \fi
  \caption{A PTA for $S_+=\{b, aa, ab\}$.}
  \label{figptaexample}
  \vspace{-20pt}
\end{wrapfigure}
\else
\fi

Given a finite and non-empty
set of positive examples over a given alphabet $\Sigma$,
$S_+ \subseteq \Sigma^*$,
we can construct, in a non-unique way, 
a tree-shaped, incomplete 
DFA, that accepts all words in $S_+$, and rejects all others. 
Such a DFA is called a {\em prefix tree acceptor}~\cite{delaHiguera2010}
(PTA) for $S_+$.
For example, a PTA for $S_+=\{b, aa, ab\}$ is shown in Figure~\ref{figptaexample}.

\ifdefined \arxiv
\else
\begin{wrapfigure}{l}{.35\textwidth}
	\vspace{-22pt}
    \centering
    \ifdefined\pdfonlyfigs
	\includegraphics[]{figures/pta_o.pdf}
	\else
	\resizebox{0.31\columnwidth}{!}{\input{figures/pta_t.pdf_tex}}
    \fi
  \caption{A PTA for $S_+=\{b, aa, ab\}$.}
  \label{figptaexample}
  \vspace{-25pt}
\end{wrapfigure}
\fi

We extend the concept of PTA to Moore machines.
Suppose that we have a set $S_{IO}$ of Moore $(I,O)$-examples.
Let $N= \left \lceil{\log_2|O|}\right \rceil$ be the number of bits necessary to represent an element of $O$. Then, given a function $f$ that maps elements of $O$ to bit tuples of length $N$, we can map $S_{IO}$ to $N$ pairs of positive and negative example sets, $\{(S_{1+}, S_{1-})$, $(S_{2+}, S_{2-})$, $\cdots$, $(S_{N+}, S_{N-})\}$. 
In particular, for each pair $(w, y) \in S_{IO}$, if the $i$-th element of $f(y)$ is $1$, then $S_{i+}$ should contain $w$ and $S_{i-}$ should not. 
Similarly, if the $i$-th element of $f(y)$ is $0$, then $S_{i-}$ should contain $w$ and $S_{i+}$ should not.

\begin{figure}[t]
  \begin{subfigure}[t]{.49\columnwidth}
    \centering
    \ifdefined\pdfonlyfigs
	\includegraphics[]{figures/ptap1_o.pdf}
	\else
	\resizebox{0.62\textwidth}{!}{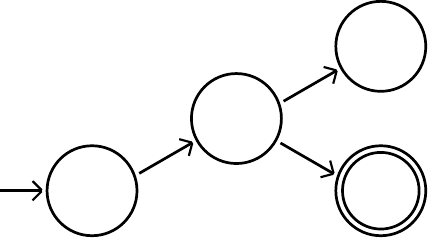}
    \fi
    \caption{The PTA for $S_{1+}=\{ab\}$.}
    \label{figptmrmpart1}
  \end{subfigure}%
  \begin{subfigure}[t]{.49\columnwidth}
    \centering
    \ifdefined\pdfonlyfigs
	\includegraphics[]{figures/ptap2_o.pdf}
	\else
	\resizebox{0.62\textwidth}{!}{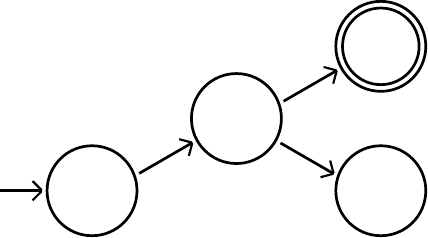}
    \fi
    \caption{The PTA for $S_{2+}=\{aa\}$.}
    \label{figptmrmpart2}
  \end{subfigure}
  \caption{A PTAP for $S_{IO}=\{(b,0), (aa,1), (ab,2)\}$,
\label{figptmrmexample}
with $I=\{a,b\}$, $O=\{0,1,2\}$, and
$f=\{0\mapsto (0,0), 1\mapsto (0,1), 2\mapsto (1,0)\}$.
The positive and negative example sets are:
$S_{1+}=\{ab\}$, $S_{1-}=\{b, aa\}$, $S_{2+}=\{aa\}$, $S_{2-}=\{b, ab\}$.}
\end{figure}

We can subsequently construct
a {\em prefix tree acceptor product} (PTAP), 
which is
a collection of $N$ PTAs, one for each positive example set, $S_{i+}$, for $i=1, \cdots, N$.
An example of a PTAP consisting of two PTAs is given in Figure~\ref{figptmrmexample}.

\section{Characteristic samples}

An important concept in automata learning theory is that of a {\em characteristic sample}~\cite{delaHiguera2010}.
A characteristic sample for a DFA is a set of words that captures all information about that automaton's set of states and behavior. In this paper we extend the concept of characteristic sample to Moore machines.

\subsection{Characteristic samples for Moore machines}
\label{sec_cs}
Let $M=(I,O,Q,q_0,\delta,\lambda)$ be a minimal Moore machine.
Let $<$ denote a total order on input words, i.e., on $I^*$, such that
$w<w'$ iff either $|w|<|w'|$, or $|w|=|w'|$ but $w$ comes before $w'$ in
lexicographic order ($|w|$ denotes the length of a word $w$).
For example, $b<aa$ and $aaa<aba$.

Given a state $q\in Q$, we
define the {\em shortest prefix} of $q$ as the shortest input word which can be
used to reach $q$:
$$S_P(q) = min_< \{w \in I^* \ |\  \delta^*(q_0, w) = q\}.$$

Notice that $M$ is minimal, which implies that all its states are reachable
(otherwise we could remove unreachable states).
Therefore, $S_P(q)$ is well-defined for every state $q$ of $M$.

Next, we define the set of {\em shortest prefixes of $M$}, denoted $S_P(M)$, as:
$$S_P(M) = \{S_P(q) \ |\  q \in Q \}$$

We can now define the {\em nucleus} of $M$
which contains the empty word and all one-letter extensions of words in
$S_P(M)$:
$$N_L(M) = \{\epsilon\} \cup \{ w \cdot a \ |\ w \in S_P(M),\ a \in I \}.$$

We also define the {\em minimum distinguishing suffix} for two different
 states $q_u$ and $q_v$ of $M$, as follows:
$$ M_D(q_u, q_v) = min_< \{ w \in I^* \ |\ \lambda^*(q_u, w) \neq \lambda^*(q_v, w) \}.$$
$M_D(q_u, q_v)$ is guaranteed to exist for any two states $q_u, q_v$ because
$M$ is minimal.

Let $W$ be a set of input words, $W\subseteq I^*$. $Pref(W)$ denotes the set
of all prefixes of all words in $W$:
$$Pref(W) = \{ x \in I^*\mid\exists w\in W, y \in I^*: x \cdot y = w \}.$$

\begin{definition}
\label{defcharactsample}
Let $S_{IO}$ be a set of Moore (I,O)-traces, and let $S_I$ be the corresponding
set of input words: $S_I = \{\rho_I\in I^*\mid (\rho_I, \rho_O) \in S_{IO}\}$.
$S_{IO}$ is a {\em characteristic sample} for a Moore machine $M$ iff:
\begin{enumerate}
\item $N_L(M) \subseteq Pref(S_I)$. \label{cond1}
\item \label{cond2}
$\forall u \in S_P(M) : \forall v \in N_L(M) : \forall w \in I^* : $
$$\delta^*(q_0, u) \neq \delta^*(q_0, v) \land w = M_D(\delta^*(q_0, u), \delta^*(q_0, v)) \Rightarrow \{ u \cdot w, v \cdot w\} \subseteq Pref(S_I).$$
\end{enumerate}
\end{definition}

For example, consider the Moore machine $M_1$ from Figure~\ref{figFSMexamples}.
We have: $S_P(q_0)=\epsilon$, $S_P(q_1)=x_2$, $S_P(M_1) = \{ \epsilon,x_2 \}$,
and $N_L(M_1) = \{ \epsilon,x_1,x_2,x_2 x_1,x_2 x_2 \}$.
The following set is a characteristic sample for $M_1$:
$$ S_{IO} = \{\ (x_1,\ y_1y_1),\ (x_2x_1,\ y_1y_2y_1),\ (x_2x_2,\ y_1y_2y_2)\ \}.$$

\begin{figure}[t]
  \begin{subfigure}[t]{.49\columnwidth}
    \centering
    \ifdefined\pdfonlyfigs
	\includegraphics[]{figures/white1_o.pdf}
	\else
	\resizebox{0.62\textwidth}{!}{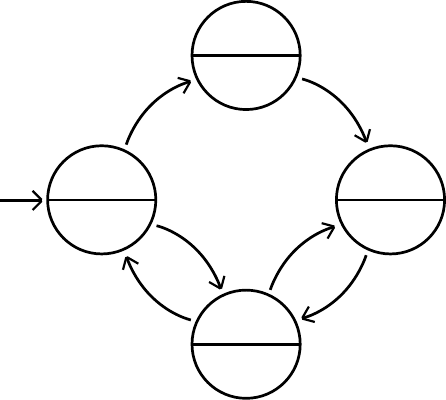}
    \fi

    \caption{Target minimal Moore machine.}
    \label{figwhiteboardex1}
  \end{subfigure}%
  \quad
  \begin{subfigure}[t]{.49\columnwidth}
    \centering
    \ifdefined\pdfonlyfigs
	\includegraphics[]{figures/white2_o.pdf}
	\else
	\resizebox{0.62\textwidth}{!}{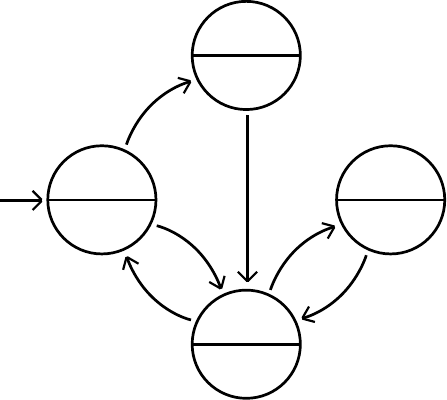}
    \fi

    \caption{Moore machine learned by our MooreMI algorithm if we use a set of traces that does not satisfy Condition~\ref{cond2} of Definition~\ref{defcharactsample}. 
    }
    \label{figwhiteboardex2}
  \end{subfigure}
  \caption{Example illustrating the need for Condition~\ref{cond2} of Definition~\ref{defcharactsample}.}
  \label{figwhiteboardexs}
\end{figure}

While it is intuitive that a characteristic sample should contain input words that in a sense {\em cover} all states and transitions of $M$ (Condition~\ref{cond1} of Definition~\ref{defcharactsample}), it may not be obvious why Condition~\ref{cond2} of Definition~\ref{defcharactsample} is necessary. This becomes clear if we look at machines having the same output on several states.
For example,
consider the Moore machine $M$ in Figure~\ref{figwhiteboardex1}. 
The set of $(I,O)$-traces $S_{IO}^1 = \{(aa, 020),$ $(ba, 012),$ $(bb, 012),$ $(aba, 0222),$ $(abb, 0222)\}$ satisfies Condition~\ref{cond1} but not Condition~\ref{cond2} (because $S_P(q_2) = a$, $ba \in N_L(M)$, $\delta^*(q_0, ba) = q_3$, $M_D(q_2, q_3) = a$, but no input word in $S_{IO}^1$ has $baa$ as a prefix), and therefore is not a characteristic sample of the machine
of Figure~\ref{figwhiteboardex1}.
If we use $S_{IO}^1$ to learn a Moore machine, we obtain the machine in Figure~\ref{figwhiteboardex2} (this machine was produced by our MooreMI algorithm,
described in Section~\ref{secPTMRPNI}).
Clearly, the two machines of Figure~\ref{figwhiteboardexs} are not equivalent.
For instance, the input word $baa$ results in different outputs when fed to the two machines. The reason why the learning algorithm produces the wrong machine is that the set $S_{IO}^1$ does not contain enough information to clearly distinguish between states $q_2$ and $q_3$.

\begin{sloppypar}
Instead, consider the set $S_{IO}^2 = \{(aa, 020),$ $(baa, 0122),$ $(bba, 0122),$ $(abaa, 02220),$ $(abba, 02220)\}$. $S_{IO}^2$ satisfies both Conditions~\ref{cond1} and~\ref{cond2}, and therefore is a characteristic sample.
Given $S_{IO}^2$ as input, our MooreMI algorithm is able to learn the correct machine, i.e., the machine of Figure~\ref{figwhiteboardex1}. In this case, the minimum distinguishing suffix of states $q_2$ and $q_3$ is simply the letter $a$, since $\delta(q_2, a) = q_0$, $\delta(q_3, a) = q_2$ and $\lambda(q_0) = 0 \neq 2 = \lambda(q_2)$. Notice that $S_{IO}^2$ can be constructed from $S_{IO}^1$ by extending with the letter $a$ the input words of the latter that land on $q_2$ or $q_3$.
\end{sloppypar}

The intuition, then, behind Condition~\ref{cond2} is that states in $M$ that have the same outputs cannot be distinguished by just those (outputs); additional suffixes that differentiate them are required.

\subsection{Computation, minimality, size, and other properties of characteristic samples}

It is easy to see that adding more traces to a characteristic sample 
preserves the characteristic sample property, i.e., if $S_{IO}$ is a
characteristic sample for a Moore machine $M$ and $S_{IO}'\supseteq S_{IO}$, 
then $S_{IO}'$ is also a characteristic sample for $M$. 
Also, arbitrarily extending the input word of an existing $(I,O)$-trace in
$S_{IO}$ and accordingly extending the corresponding output word, again yields a new characteristic sample for $M$. The questions are raised, then, whether there exist characteristic samples that are minimal in some sense, how many elements they consist of, what are the lengths of their elements, and how can we construct them.

\ifdefined \arxiv
In the following, 
\else
In~\cite{LearningMooreArxiv2016},
\fi
we outline a simple procedure that, given a minimal Moore machine $M$, returns a characteristic sample $S_{IO}$ that is minimal in the sense that removing any $(I,O)$-trace from it or dropping any number of letters at the end of an input word in it (and accordingly adjusting the corresponding output word) will result in a set that is not a characteristic sample. By doing so, we also constructively establish the existence of at least one characteristic sample for any minimal Moore machine $M$.
\ifdefined \arxiv

Let $M=(I,O,Q,q_0,\delta,\lambda)$ be a minimal Moore machine, $S_I$ an initially empty set of input words and $S_{IO}$ the set of $(I,O)$-traces formed by the elements of $S_I$ and the corresponding output words. We compute $S_P(M)$ and $N_L(M)$, and add the elements of the latter to $S_I$. Then, for each pair of words $(u, v) \in S_P(M) \times N_L(M)$ leading to {\em different} states $q_u = \delta^*(q_0, u), \ q_v = \delta^*(q_0, v)$, we compute $M_D(q_u, q_v)$ and add it to $S_I$. Now, $S_{IO}$ already is a characteristic sample. However, it may contain redundant elements that can safely be removed. We can do this by simply considering each element of $S_I$ and removing it if it is a prefix of another element (this step can be sped up by choosing an appropriate data structure to represent $S_I$, e.g. using a trie, we would simply just keep the words represented by the leaf nodes). Note that since the prefix relation on words is a partial order, and therefore transitive, the order in which we remove the redundant elements does not affect the final result. It is easy to see now that, after this step, (1) no element of $S_I$ is the prefix of another, (2) $S_{IO}$ is still a characteristic sample, and (3) removing any element from $S_I$ or dropping any number of letters at the and of it, will result in $S_{IO}$ not being a characteristic sample.

By definition, there is a $1-1$ correspondence between the elements of $S_P(M)$ and the states of $M$. Therefore, $|S_P(M)| = |Q|$. It follows that $|N_L(M)| \leq |S_P(M)|\cdot|I| + 1 = |Q|\cdot|I| + 1$ and, consequently, $|S_{IO}| = |S_I| \leq |N_L(M)| + |S_P(M)| \cdot |N_L(M)| = (|Q|\cdot|I| + 1)\cdot(|Q| + 1) $. In other words, the size of $S_{IO}$ is $O(|Q|^2|I|)$.

We now provide bounds on the lengths of the elements of $S_{IO}$. The lengths of shortest prefixes are bounded by the longest non-looping path in $M$, which in turn is bounded by $|Q|$. It follows that the nucleus element lengths are bounded by $|Q| + 1$. Let now $q_u$ and $q_v$ be different states of $M$ and consider $M_1 = (I, O, Q, q_u, \delta, \lambda)$ and $M_2 = (I, O, Q, q_v, \delta, \lambda)$, i.e. $M_1$ and $M_2$ have $q_u$ and $q_v$ as initial states, respectively, but are otherwise identical to $M$. Finding a (minimum) distinguishing suffix of $q_u$ and $q_v$ is now reduced to finding a (minimum) input word that leads to different output words when transduced by $M_1$ and $M_2$. To find such a word, we first construct a DFA $A = (I, Q \times Q, (q_u, q_v), \delta_A, F)$, where $\forall (q_1, q_2) \in Q \times Q : \forall a \in I : \delta_A( (q_1, q_2), a) = (\delta(q_1, a), \delta(q_2, a))$ and $F = \{ (q_1, q_2) \in Q \times Q~|~\lambda(q_1) \neq \lambda(q_2) \}$. A word accepted by this DFA is a distinguishing suffix of $q_u$ and $q_v$, and it is easy to see that we only need to test words of length up to $|Q \times Q|$ in order to find one. We can conclude from the above that the sum of lengths of elements in $S_{IO}$ is $O(|Q|^4|I|)$.
\else

Space limitations prevent us from including the description of the procedure
in this paper: it can be found in~\cite{LearningMooreArxiv2016}, 
together with an analysis of the procedure in order to determine the ``size''
of the characteristic sample $S_{IO}$. It seems reasonable to measure this
size as the sum of the lengths of all elements in $S_{IO}$. As it turns out,
this sum is $O(|Q|^4|I|)$.
\fi

\section{Learning Moore machines from Input-Output Traces}

\subsection{Problem definition}
\label{sec_problem}

The problem of learning Moore machines from input-output traces (LMoMIO)
is defined as follows.
Given an input alphabet $I$, an output alphabet $O$, and a set $\trainset{}$ 
 of Moore $(I,O)$-traces, called the {\em training set}, we want to
synthesize automatically a deterministic, complete Moore machine $M=(I,O,Q,q_0,\delta,\lambda)$, such that $M$ is consistent with $\trainset{}$, i.e.,
$\forall\ (\rho_I, \rho_O) \in \trainset{} : \lambda^*(\rho_I)=\rho_O$.
($\trainset{}$ is assumed to be itself consistent, in the sense it
does not contain two different pairs with the same input word.)

In addition to consistency, we would like to evaluate our learning technique w.r.t.  various {\em performance} criteria, including:
\begin{itemize}
\item {\em Size} of $M$, in terms of number of states.
Note that, contrary to the
{\em exact identification} problem~\cite{DBLP:journals/iandc/Gold78},
we do {\em not require} $M$ to be the smallest (in terms of number of states)
machine consistent with $\trainset{}$. 
\item {\em Accuracy} of $M$, which, informally speaking, is a measure of
how well $M$ performs on a set of traces, $\testset{}$, {\em different} from the training set. $\testset{}$ is called the {\em test set}. Accuracy is a standard criterion in machine learning. 
\item {\em Complexity} (e.g., running time) of the learning algorithm itself.
\end{itemize}

In the rest of this paper, we present three learning algorithms which solve the LMoMIO problem, and evaluate them w.r.t. the above criteria. Complexity of the algorithm and size of the learned machine are standard notions.
Accuracy is standard in machine learning topics such as classification,
but not in automata learning.
Thus, we elaborate on this concept next. 

There are more than one ways to measure the accuracy of a learned Moore machine $M$ against a test set $\testset{}$. We call an {\em accuracy evaluation policy} (AEP) any function that, given a Moore $(I,O)$-trace $(\rho_I, \rho_O)$
and a Moore machine $M = (I, O, Q, q_0, \delta, \lambda)$, will return a real number in $[0, 1]$. We will call that number the accuracy of $M$ on $(\rho_I, \rho_O)$. In this paper, we use three AEPs which we call {\em strong}, {\em medium}, and {\em weak}, defined below. 
Let $(\rho_I, \rho_O) = (x_1 x_2 \cdots x_n,\ y_0 y_1 \cdots y_n)$
and $z_0 z_1 \cdots z_n = \lambda^*(q_0, \rho_I)$.

\begin{itemize}
\item {\em Strong}:
if $\lambda^*(q_0, \rho_I) = \rho_O$ then 1 else 0. 
\item {\em Medium}: $\frac{1}{n+1}\cdot |\{ i\mid y_0 y_1 \cdots y_i = z_0 z_1 \cdots z_i\}|$.

\item {\em Weak}: $\frac{1}{n+1} \cdot |\{ i ~|~ y_i = z_i \}|$.

\end{itemize}

The strong AEP says that the output of the learned machine $M$ must be
identical to the output in the test set.
The medium AEP returns the proportion of the largest output prefix that matches.
The weak AEP returns the number of output symbols that match.
For example, if the correct output is $0012$ and $M$ returns $0022$ then the 
strong accuracy is 0, the medium accuracy is $\frac{2}{4}$, and the weak accuracy is $\frac{3}{4}$. 
Ideally, we want the learned machine to achieve a high accuracy with respect to
the strong AEP. However, the medium and weak AEPs are also useful, because they
allow to distinguish, say, a machine which is ``almost right'' (i.e., outputs
the right sequence except for a few symbols) from a machine which is always or
almost always wrong.

Given an accuracy evaluation policy $f$ and a test set $\testset{}$, we define the accuracy of $M$ on $\testset{}$ as the averaged accuracy of $M$ over all traces in $\testset{}$, i.e.,
$$
\frac{\sum_{(\rho_I, \rho_O)\in \testset{}} f((\rho_I, \rho_O), M)}{|\testset{}|}.
$$

It is often the case that the test set $\testset{}$ contains traces generated by a ``black box'', for which we are trying to learn a model.
Suppose this black box corresponds to an unknown machine $M_?$.
Then, ideally, we would like the learned machine $M$ to be equivalent to
$M_?$. In that case, no matter what test set is generated by $M_?$, the
learned machine $M$ will always achieve 100\% accuracy. 
Of course, achieving this ideal depends on the training set: if the latter
is ``poor'' then it does not contain enough information to identify the
original machine $M_?$.
A standard requirement in automata learning theory states that when the 
training set is a characteristic sample of $M_?$, then the learning algorithm
should be able to produce a machine which is equivalent to $M_?$.
We call this the {\em characteristic sample requirement} (CSR).
CSR is important, as it ensures {\em identification in the limit},
a key concept in automata learning theory~\cite{DBLP:journals/iandc/Gold67}.
In what follows, we show that among the algorithms that will be
presented in~\S\ref{sec_algos}, only MooreMI satisfies CSR.

\begin{sloppypar}
Before proceeding, we remark that a given Moore $(I,O)$-trace
$(\rho_I,\rho_O)=(x_1 x_2 \cdots x_n,\; y_0 y_1 \cdots y_n)$
can be represented as a set of $n+1$ Moore $(I,O)$-examples, specifically $\{(\epsilon, y_0)$, $(x_1, y_1)$, $(x_1 x_2, y_2)$, $\cdots$, $(x_1 x_2 \cdots x_n, y_n)\}$.
Because of this observation, in all approaches discussed below, there is a preprocessing step to convert the training set, first into an equivalent set of Moore $(I,O)$-examples, and second, into an equivalent set of 
$N$ pairs of positive and negative example sets
(the latter conversion was described in \S\ref{sec_ptas}).
\end{sloppypar}

\subsection{Algorithms to solve the LMoMIO problem}
\label{sec_algos}

\subsubsection{The PTAP algorithm}
This algorithm is a rather straightforward one. The set of Moore $(I,O)$-examples obtained after the preprocessing step described above is used to construct a PTAP, as described in \S\ref{sec_ptas}.
Recall that a PTAP is a collection of $N$ PTAs having the same state-transition structure.
The synchronous product of these $N$ PTAs is then formed, {\em completed}, and returned as the result of the algorithm. 
Note that a PTA is a special case of an NFA: the PTA is deterministic, but it is generally incomplete. The synchronous product of PTAs is therefore the same as 
the synchronous product of NFAs. The product of PTAs is deterministic, but also generally incomplete, and therefore needs to be completed in order to yield a complete DFA.
{\em Completion} in this case consists in adding self-loops to states that are missing outgoing transitions for some input symbols. The added self-loops are labeled with the missing input symbols.

Although the PTAP algorithm is relatively easy to implement and runs efficiently, it has several drawbacks. First, since no state minimization is attempted, the resulting Moore machine can 
be unnecessarily large.
Second, and most importantly, the produced machines generally have poor accuracy  
since completion is done in a trivial manner.

\subsubsection{The PRPNI algorithm}

Again, consider the $N$ pairs of positive and negative example sets obtained
after the preprocessing step.
The PRPNI algorithm starts by executing the RPNI DFA learning algorithm~\cite{Oncina92identifyingregular} on each pair, thus obtaining $N$ learned DFAs. Then, the synchronous product of these DFAs is formed, {\em completed}, and returned as the algorithm result.
As in the case of the PTAP algorithm, the synchronous product of the DFAs 
in the PRPNI algorithm is deterministic but generally not complete.

The completion step of the PRPNI algorithm is more intricate than the completion step of the PTAP algorithm.
The reason is that the synchronous product of the learned DFAs may contain
reachable states whose bit encoding does not correspond to any valid output in $O$. For example, suppose $O=\{0,1,2\}$, so that we need $2$ bits to encode it, and thus $N=2$ and we use RPNI to learn $2$ DFAs. Suppose the encoding is $0\mapsto 00, 1\mapsto 01, 2\mapsto 10$. This means that the code $11$ does not correspond to any valid output in $O$. However, it can still be the case that in the product of the two DFAs there is a reachable state with the output $11$, i.e., where both DFAs are in an accepting state. Note that this problem does not arise in the PTAP algorithm, because all PTAs there are guaranteed to have the same state-transition structure, which is also the structure of their synchronous product.

To solve this invalid-code problem, we assign all invalid codes to an arbitrary valid output. In our implementation, we use the lexicographic minimum. In the above example, the code $11$ will be assigned to the output $0$.

Compared to the PTAP algorithm, the PRPNI algorithm has the
advantage of being able to learn a minimal Moore machine when provided with enough $(I,O)$-traces. However, it can also perform worse in terms of both running time and size (number of states) of the learned machine, due to potential state explosion while forming the DFA product. The PTAP algorithm does not have this problem because, as explained above, the structure, and therefore also the number of states, of the product is identical to those of the component PTAs.

\subsubsection{The MooreMI algorithm}
\label{secPTMRPNI}

As we saw above, both the PTAP and PRPNI algorithms have several drawbacks.
In this section we propose a novel algorithm, called, MooreMI,
 which remedies these.  
Moreover, we shall prove that MooreMI satisfies CSR.

The MooreMI algorithm begins by building a PTAP represented as $N$ PTAs, as in the PTAP algorithm. Then, a merging phase follows, where a merge operation is accepted only if all resulting DFAs are consistent with their respective negative example sets. In addition, a merge operation is either performed on all DFAs at once or not at all. Finally, the synchronous product of the $N$ learned DFAs is formed, completed by adding self loops for any missing input symbols, as in the PTAP algorithm, and returned.
\ifdefined \arxiv
The pseudocode of the algorithm is given below.
\else
The pseudocode for the MooreMI algorithm can be found in~\cite{LearningMooreArxiv2016}.
Space limitations prevent us from including it in this paper.
\fi

\ifdefined \arxiv
The main \texttt{MooreMI} procedure calls the $merge$ function as a subroutine.
$merge$ computes the result of merging the given red and blue states of the given DFA component. It also performs additional potentially necessary state merges to preserve determinism.

\begin{lstlisting}[mathescape]
def MooreMI(trace_set, $\Sigma_I$, $\Sigma_O$):

  (list_of_pos_example_sets, 
   list_of_neg_example_sets, 
   bits_to_output_func) 
     := $preprocess\_moore\_traces$(trace_set)
    
  N := $ceil$( $log_2$( |$\Sigma_O$| ) )
    
  DFA_list := $build\_prefix\_tree\_acceptor\_product$(
    	    list_of_pos_example_sets, $\Sigma_I$, $\Sigma_O$)
    
  red = { $q_\epsilon$ }
  blue = { $q_a$ for $a$ in $\Sigma_I$ } $\cap$ DFA_list[0].Q
    
  while blue $\neq~\emptyset$:
    
    q_blue = $pick\_next$(blue)
    blue := blue - {q_blue}
        
    merge_accepted := false
        
    for q_red $\in$ red:
            
      for i $\in$ {0, ..., N - 1}:
        new_DFA_list[i] :=
          $merge$(DFA_list[i], q_red, q_blue) 
            
      if $\forall$ i $\in$ {0, ..., N - 1}:
             $is\_consistent$(
                  new_DFA_list[i], 
                  list_of_neg_example_sets[i]):
          merge_accepted := true
          break
                
    if merge_accepted:
      DFA_list := new_DFA_list
      blue := blue $\cup$ ( { one-letter 
          successors of red states } 
          $\cap$ DFA_list[0].Q )
    else:
      red := red $\cup$ {q_blue}
      blue := blue $\cup$ ( { one-letter 
          successors of q_blue } 
          $\cap$ DFA_list[0].Q )

  return $product$(
    DFA_list, 
    bits_to_output_func).$make\_complete$()

def $merge$(DFA, q_red, q_blue):
   
  q_u := $unique\_parent\_of$(q_blue)
  a_u := $unique\_input\_from\_to$(q_u, q_blue)

  DFA.$\delta$(q_u, a_u) := q_red    

  merge_stack := [(q_red, q_blue)]    
    
  while merge_stack $\neq$ []:
    
    (q_1, q_2) := $pop$(merge_stack)
        
    if q_1 = q_2 : continue
        
    if (q_1, q_2) $\neq$ (q_red, q_blue) 
        and q_2 < q_1:
      q_1, q_2 := q_2, q_1 		

    DFA.$Q$ := DFA.$Q$ - {q_2}
        
    if q_2 $\in$ DFA.$F$:
      DFA.$F$ := DFA.$F$ $\cup$ {q_1}
            
    for a $\in$ DFA.$\Sigma$:
      if $is\_defined$(DFA.$\delta$(q_2, a)):
        if $is\_defined$(DFA.$\delta$(q_1, a)):
                    
          $push$(merge_stack, 
                DFA.$\delta$(q_1, a), 
                DFA.$\delta$(q_2, a)))                    
        else:
          DFA.$\delta$(q_1, a) := DFA.$\delta$(q_2, a)
                    
  return DFA
\end{lstlisting}

After the initial preprocessing step (line 6), the algorithm builds a prefix tree acceptor product (line 10) and then repeatedly attempts to merge states in it, in a specific order (line 16). While not appearing in the original RPNI algorithm, the convention of marking states as {\em red} or {\em blue} was introduced later in \cite{DBLP:conf/icgi/LangPP98}. States marked as red represent states that have been processed and will be part of the resulting machine. States marked as blue are immediate successors of red states and represent states that are currently being processed. Initially, the set of red states only contains the initial state $q_\epsilon$, and the set of blue states contains the one-letter successors of $q_\epsilon$ (lines 13, 14). Unmarked states will eventually become blue (lines 38, 43), and then either merged with red ones (lines 27, 36) or become red states themselves (line 42).

Most of the auxiliary functions whose implementations are not shown in the pseudocode have self explanatory names. For instance, the $push$ and $pop$ functions push and pop, respectively, elements to / from a stack, and the functions in lines 53, 54 compute the unique parent of and corresponding input symbol leading to the given blue state (uniqueness of both is guaranteed by the tree-shaped nature of the initial PTA). The function $pick\_next$, however, deserves some additional explanation. Notice first that after the prefix tree acceptor product is constructed and before the merging phase of the algorithm begins, each state can reached by a unique input word which is used to identify that state. For example, the state reached by the word $abba$ is referred to as state $q_{abba}$. The word used to identify a state may change during merging operations. The total order on words $<$ defined in \S\ref{sec_cs} can now naturally be extended on states of the learned machine as follows: $q_u < q_v \iff u < v$, in which case we say that $q_u$ is smaller than $q_v$. The $pick\_next$ function simply returns the smallest state of the blue set, according to the order we just defined. 
\fi

MooreMI is able to learn minimal Moore machines,
while avoiding the state explosion and invalid code issues of PRPNI. To see this, notice first that, at every point of the algorithm, the $N$ learned DFAs are identical in terms of states and transitions, modulo the marking of their states as final. Indeed, this invariant holds by construction for the $N$ initial prefix tree acceptors, and the additional merge constraints make sure it is maintained throughout the algorithm. Therefore, the product formed at the end of the algorithm can be obtained by simply ``overlaying'' the $N$ DFAs on top of one another, as in the PTAP approach, which implies no state explosion. The absence of invalid output codes is also easy to see. Invalid codes generally are results of problematic state tuples in the DFA product, that cannot appear in MooreMI due to the additional merge constraints. Indeed, if a state tuple occurs in the final product, it must also occur in the initial prefix tree acceptor product, and if it occurs there, its code cannot be invalid.

\subsection{Properties of the algorithms}

All three algorithms described above satisfy consistency w.r.t. the input training set. For PTAP and PRPNI, this is a direct consequence of the properties of PTAs, of the basic RPNI algorithm, and of the synchronous product. 
The proof for MooreMI is somewhat more involved, therefore the result
for MooreMI is stated as a
\ifdefined \arxiv
theorem:
\else
theorem (proofs can be found in~\cite{LearningMooreArxiv2016}):
\fi

\begin{theorem}[Consistency]
\label{thm:theoremone}
The output of the MooreMI algorithm is a complete Moore machine, consistent with the training set. Formally, let $S_{IO}$ be the set of Moore $(I,O)$-traces used as input for the algorithm, and let $M=(I,O,Q,q_0,\delta,\lambda)$ be the resulting Moore machine. Then, $\delta$ and $\lambda$ are total functions and $\forall\ (\rho_I, \rho_O) \in S_{IO}\ :\ \lambda^*(q_0, \rho_I)=\rho_O$.
\end{theorem}

\ifdefined \arxiv
\begin{proof}%[Theorem \ref{thm:theoremone}]
\input{proofs/theorem1.tex}
\end{proof}
\fi

We now show that MooreMI satisfies the characteristic sample requirement, i.e.,
if it is fed with a characteristic sample for a machine $M$, then it learns
a machine equivalent to $M$. If $M$ is minimal then the learned machine will
in fact be isomorphic to $M$.
\ifdefined \arxiv
We first introduce some auxiliary definitions and notation, and make
some observations which are important for the proof of the result.

Let $M =(I, O, Q_m, q_{0\_m}, \delta_m, \lambda_m)$ be the minimal Moore machine from which we derive a characteristic sample,
then given as input to the MooreMI algorithm.
Let $M_A =(I, O, Q_A, q_\epsilon, \delta_A, \lambda_A)$ be the machine produced
by the algorithm.
We will use plain $Q$ and $\delta$ to denote the state set and possibly partial transition function of the learned machine in an intermediate step of the algorithm. 
\fi

\ifdefined \arxiv

\input{textwithlemmas}

\fi

\begin{theorem}[Characteristic sample requirement]
\label{thm:theoremtwo}
If the input to MooreMI is a characteristic sample of a minimal Moore machine $M$, then the algorithm returns a machine $M_A$ that is isomorphic to $M$.
\end{theorem}

\ifdefined \arxiv
\begin{proof}%[Theorem \ref{thm:theoremtwo}]
\input{proofs/theorem2.tex}
\end{proof}
\fi

Finally, we show that the MooreMI algorithm achieves identification in the limit.

\begin{sloppypar}
\begin{theorem}[Identification in the limit]
\label{thm:theoremthree}
Let $M=(I, O, Q, q_{0}, \delta, \lambda)$ be a minimal Moore machine.
Let $(\rho_I^1,\rho_O^1), (\rho_I^2,\rho_O^2), \cdots $ be an infinite
sequence of $(I,O)$-traces generated from $M$, such that
$\forall \rho\in I^* : \exists i : \rho = \rho_I^i$ (i.e., every input
word appears at least once in the sequence).
Then there exists index $k$ such that for all $n\ge k$,
the MooreMI algorithm learns a machine
equivalent to $M$ when given as input the training set
$\{(\rho_I^1,\rho_O^1), (\rho_I^2,\rho_O^2), \cdots , (\rho_I^n,\rho_O^n)\}$.
\end{theorem}
\end{sloppypar}

\ifdefined \arxiv
\begin{sloppypar}
\begin{proof}%[Theorem \ref{thm:theoremthree} - Identification in the limit]
\input{proofs/theorem3.tex}

\end{proof}
\end{sloppypar}
\fi

\ifdefined \arxiv
\subsection{Performance optimizations}   

Compared to the pseudocode our implementation includes several optimizations.
First, to limit the amount of copying involved in performing a $merge$ operation, we perform the required state merges in-place, and at the same time record the actions needed to undo them in case the merge is not accepted.

Second, the $merge$ function needs to know the unique (due to the tree-shaped nature of PTAs) parent of the blue state passed to it as an argument. The naive way of doing this, simply iterating over the states until we reach the parent, can seriously harm performance. Instead, in our implementation, we build during PTA construction, and maintain throughout the algorithm, a mapping of states to their parents, and consult this when needed. 

Third, in the negative examples consistency test, many of the acceptance checks involved are redundant. For example, suppose that starting from the initial state it is only possible to reach red states (i.e. not blue or unmarked ones) within $n$ steps (transitions). Then, there is no need to include negative examples of length less than $n$ in the consistency test. Our implementation optimizes such cases by integrating the consistency test with the merge operation. In particular, we construct the initial PTAs based not only on positive but also on negative examples, and mark states not only as accepting but also as rejecting when appropriate, as described in \cite{Coste1998}. Then, during the merge operation, if an attempt to merge an accepting state with a rejecting one occurs, the merge is rejected. 

\fi

\subsection{Complexity analysis}

\ifdefined \arxiv
In order to build a prefix tree acceptor we need to consider
all prefixes of words in the set of positive examples $S_+$. This yields a complexity of $O(\sum_{w \in S_+} |w|)$, where $|w|$ indicates the length of the word $w$. A prefix tree acceptor product is represented by $N$ prefix tree acceptors that have the same state-transition structure, where $N$ is the number of bits required to represent an output letter. Therefore, constructing a prefix tree acceptor product having $2^{N-1} < |O| \leq 2^N$ distinct output symbols, requires $O(N\cdot \sum_{w \in S_+^{all}} |w|)$ work, where $S_+^{all}$ denotes the union of the $N$ positive example sets, $S_{i+}$ (we need to consider all for each PTA, because we want the PTAs to have the same state-transition structure). 

During the main loop of the basic RPNI algorithm, at most $|Q_{PTA}|^2$ merge operations are attempted, where $Q_{PTA}$ denotes the set of states in the PTA. Each merge operation (including all additional state merges required to maintain determinism) requires $O(|Q_{PTA}|)$ work. After every merge operation, a compatibility check is performed to determine whether it should be accepted or not, requiring $O(\sum_{w \in S_-} |w|)$ work. Bearing in mind that $|Q_{PTA}|$ is bounded by $\sum_{w \in S_+} |w|$, all this amounts for a total work in the order of $O((\sum_{w \in S_+} |w|)^2\cdot (\sum_{w \in S_+} |w| + \sum_{w \in S_-} |w|))$. 

In the PRPNI algorithm, the basic RPNI loop is repeated $N$ times in sequence, which amounts for a total complexity of $O(\sum_{i=1}^{N} (\sum_{w \in S_{i+}} |w|)^2\cdot (\sum_{w \in S_{i+}} |w| + \sum_{w \in S_{i-}} |w|))$. In the MooreMI approach, $N$ DFAs are learned in parallel, and the total work done is $O(N \cdot (\sum_{w \in S_+^{all}} |w|)^2\cdot (\sum_{w \in S_+^{all}} |w| + \sum_{w \in S_-^{all}} |w|))$, where $S_-^{all}$, similarly to $S_+^{all}$, denotes the union of the $N$ negative example sets, $S_{i-}$. Note here that since the sets $S_{i+}$ (resp. $S_{i-}$) are not disjoint in general, $\sum_{w \in S_+^{all}} |w|$ (resp. $\sum_{w \in S_-^{all}} |w|$) is bounded by $\sum_{i=1}^{N} \sum_{w \in S_{i+}} |w|$ (resp. $\sum_{i=1}^{N} \sum_{w \in S_{i-}} |w|$). 

Forming the DFA product to obtain a Moore machine requires $O(N \cdot |Q_{PTA}|)$ work for the PTAP and MooreMI algorithms, but $O(N \cdot \prod_{i=1}^{N} |Q_{PTA}^i|)$ work for the PRPNI approach. Similarly, completing the resulting Moore machine requires $O(|I| \cdot |Q_{PTA}|)$ work for the PTAP and MooreMI algorithms, and $O(|I| \cdot \prod_{i=1}^{N} |Q_{PTA}^i|)$ work for PRPNI, where $I$ is the input alphabet (which can be inferred from the training set).

Note that the above hold in the case we do not apply the final performance optimization. If we do, the terms corresponding to consistency checks ($\sum_{w \in S_{i-}} |w|$, $\sum_{w \in S_-^{all}} |w|$) are removed, and, since the prefix tree acceptors are now built using both positive and negative examples, $S_+$ and $S_+^{all}$ are replaced by $S_+ \cup S_-$ and $S_+^{all} \cup S_-^{all}$, respectively.

Summarizing the above, let
\else
Let
\fi
$I$ and $O$ be the input and output alphabets, and let $S_{IO}$ be the set of Moore $(I, O)$-traces provided as input to the learning algorithms. Let $N = \lceil log_2(|O|) \rceil$ be the number of bits required to encode the symbols in $O$. Let $S_{1+}, S_{1-}, ..., S_{N+}, S_{N-}$ be the positive and negative example sets obtained by the preprocessing step at the beginning of each algorithm.
Let $m_+ = \sum_{i=1}^{N} \sum_{w \in S_{i+}} |w|$, $m_- = \sum_{i=1}^{N} \sum_{w \in S_{i-}} |w|$, and $k=\sum_{(\rho_I, \rho_O) \in S_{IO}} |\rho_I|^2 $.
 The time required for the preprocessing step is $O(N \cdot k )$, and is the same for all three
algorithms. The time required for the rest of the phases of each algorithm is $O((N + |I|) \cdot m_+)$ for PTAP, $O((N + |I|) \cdot m_+^N + N \cdot m_+^2\cdot (m_+ + m_-))$ for PRPNI, and $O((N + |I|) \cdot m_+ + N \cdot m_+^2\cdot (m_+ + m_-) )$ for MooreMI. 
It can be seen that the complexity of MooreMI is no more than logarithmic in the number of output symbols, linear in the number of inputs, and cubic in the total length of training traces. 
This polynomial complexity does not contradict Gold's NP-hardness result \cite{DBLP:journals/iandc/Gold78}, since the problem we solve is not the exact identification problem (c.f. also Section~\ref{sec_related}).

\section{Implementation \& Experiments}

All three algorithms presented in \S\ref{sec_algos} have been implemented in Python. The source code, including random Moore machine and characteristic sample generation, learning algorithms and testing, spans roughly 2000 lines of code.
The code and experiments are available upon request.

\subsection{Experimental comparison}

We randomly generated several minimal Moore machines of sizes 50 and 150 states, and input and alphabet sizes $|I|=|O|=25$
\ifdefined \arxiv
.\footnote{The random generation procedure takes as inputs a random seed, the number of states, and the sizes of the input and output alphabets of the machine. Two intermediate steps are worth mentioning: (1) After assigning a random output to each state, we fix a random permutation of states and assign the $i$-th output to the $i$-th state. This ensures that all output symbols appear in the machine. (2) After assigning random landing states to each (state, letter) pair, we fix a random permutation of states that begins with the initial state, and add transitions with random letters from the $i$-th to the $(i+1)$-th state. This ensures that all states in the machine are reachable. Finally, a minimization algorithm is employed to make sure the generated machine is indeed minimal.}
\else
(see \cite{LearningMooreArxiv2016} for details on the random generation method).
\fi
From each such machine, we generated a characteristic sample, and ran each of the three algorithms on this characteristic sample, i.e., using it as the training set.
Then we took the learned machines generated by the algorithms, and evaluated 
these machines in terms of size (\# states) and accuracy. For accuracy,
we used a test set of size double the size of the training set.
The length of words in the test set was double the maximum training word length.

The results are shown in Table~\ref{tab50states}.
``Algo 1,2,3'' refers to PTAP, PRPNI, and MooreMI, respectively.
``Time'' refers to the average execution time of the learning algorithm, in seconds.
``States'' refers to the average number of states of the learned machines.
For accuracy, we used the three AEPs, Strong, Medium, and Weak, defined
in \S\ref{sec_problem}. 
The table is split into two tabs according to the size of the original machines mentioned above. Each row represents the average performance of an algorithm over training sets generated by 5 different Moore machines. The only exception is row 2 of the 50 states tab, 
where one of the 5 experiments timed out and the reported averages are over 4 experiments.
``Timeout'' means that the algorithm was unable to terminate within the given time limit (60 seconds) in any of the 5 experiments with 150 states.\footnote{
Note, however, that our algorithms perform better in terms of execution time than approaches that solve exact identification problems.
For example, \cite{DBLP:journals/corr/UlyantsevBS16} report experiments where
learning a Mealy machine of 18 states requires more than 29 hours. 
The majority of the execution time here is spent in proving that there exists
no machine with fewer than 18 states which is also consistent with the examples.
Since we don't require the smallest machine, our algorithms avoid this penalty.}

\begin{table}[t]
\caption{50 (resp. 150) states tab: average training set size: 1305 (resp. 4540), average input word length in training set: 3.5 (resp. 4).}
\label{tab50states}
\centering
\begin{tabular}{| c | c | c | c | c | c | c | c | c | c | c |}
\hline
 & \multicolumn{5}{c|}{50 states} & \multicolumn{5}{c|}{150 states} \\ \hline 
 \multirow{2}{*}{Algo} & \multirow{2}{*}{~Time~} & \multirow{2}{*}{~States~}  & \multicolumn{3}{c|}{Accuracy (\%)} 
   & \multirow{2}{*}{~Time~} & \multirow{2}{*}{~States~} & \multicolumn{3}{c|}{Accuracy (\%)}  \\ \cline{4-6} \cline{9-11}
 
& & & Strong & Medium & Weak & & & Strong & Medium & Weak \\ \hline
1     &    0.973   &    2113     &   0 & 32.44 & 35.39  &    8.329   &    7135     &   0 & 28.28 & 31.13  \\
2     &   12.753     &    8925    &    0 &    33.82 &    36.57  &   60     &    Timeout    &    - &    - &    -  \\
3     &     0.348    &    50     &  100 &  100 &  100  &     2.545    &    150     &  100 &  100 &  100  \\
\hline
\end{tabular}
\end{table}

As expected, MooreMI always achieves 100\% accuracy, since the input is a characteristic sample (we verified that indeed the machines learned by MooreMI are in each case equivalent to the original machine that produced the training set).
But as it can be seen from the table, neither PTAP nor PRPNI learn the correct machines, even though the training set is a characteristic sample.

The table also shows that PTAP and PRPNI generate much larger machines than the correct ones. This in turn explains why MooreMI performs better in terms of running time than the other two algorithms, which spend a lot of time completing the large number of generated states.

\subsection{Comparison with OSTIA}

\ifdefined \arxiv
\else
\begin{wrapfigure}{r}{.35\textwidth}
\vspace{0pt}
    \centering
    \ifdefined\pdfonlyfigs
	\includegraphics[]{figures/ostia_o.pdf}
	\else
	\resizebox{0.36\columnwidth}{!}{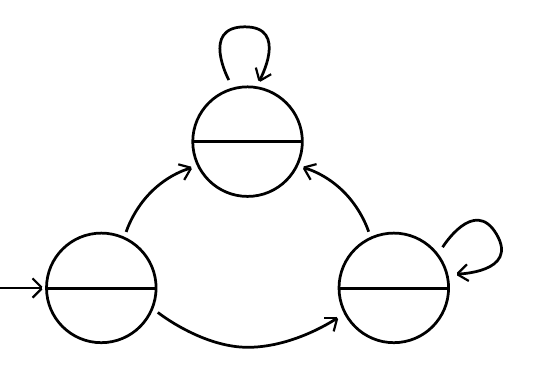}
    \fi  
  \caption{\small The transducer learned by OSTIA 
  	given a characteristic sample for
  	the Moore machine in Figure~\ref{figwhiteboardex1} as input.}
  \label{figostiaexample}
\vspace{-15pt}
\end{wrapfigure}
\fi

\ifdefined \arxiv
\begin{wrapfigure}{r}{.35\textwidth}
\vspace{-20pt}
    \centering
    \ifdefined\pdfonlyfigs
	\includegraphics[]{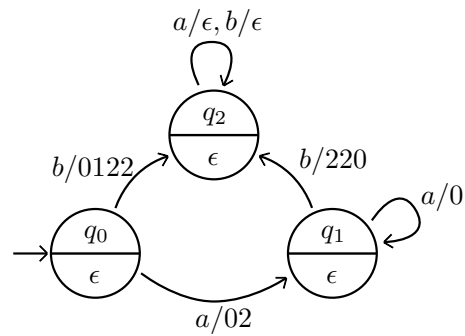}
	\else
	\resizebox{0.36\columnwidth}{!}{\input{figures/ostia_t.pdf_tex}}
    \fi  
  \caption{\small The transducer learned by OSTIA 
  	given a characteristic sample for
  	the Moore machine in Figure~\ref{figwhiteboardex1} as input.}
  \label{figostiaexample}
\vspace{-30pt}
\end{wrapfigure}

\else

\fi

OSTIA~\cite{DBLP:journals/pami/OncinaGV93} is a well-known algorithm that learns {\em onward subsequential transducers}, a class of transducers more general than Moore and Mealy machines. Then, a question arising naturally is whether it is possible to use OSTIA for learning Moore machines. In particular, we would like to know what happens when the input to OSTIA is a set of Moore (I,O)-traces: will OSTIA learn a Moore machine? 

The answer here is negative, as indicated by an experiment we performed. We constructed a characteristic sample for the Moore machine in Figure~\ref{figwhiteboardex1} and ran the OSTIA algorithm on it (we used the open source implementation described in \cite{Akram:2010:GIA:1886263.1886288}). The resulting machine is depicted in Figure~\ref{figostiaexample}. Notice that there are transitions whose corresponding outputs are words of length more than 1 (e.g.,
transition label $b/0122$), or even the empty word (output of initial state
$q_0$). We conclude that in general OSTIA cannot learn Moore machines, even when the training set is a set of Moore traces, and is also a characteristic sample.

\section{Conclusion \& Future Work}

We formalized the problem of learning Moore machines for input-output traces
and developed three algorithms to solve this problem.
We showed that the most advanced of these algorithms, MooreMI, has desirable
theoretical properties: in particular it satisfies the characteristic sample requirement and achieves identification in the limit.
We also compared the algorithms experimentally and showed that MooreMI is also
superior in practice.

Future work includes:
(1) studying learning for Mealy and other types of state machines;
(2) developing incremental versions of the learning algorithms
presented here;
(3) further implementation and experimentation;
and
(4) application of the methods presented here for learning models
of various types of black-box systems.

\bibliographystyle{abbrv}

\bibliography{arxiv-v2}

\appendix

\end{document}

%% file: figures/moore_t.pdf_tex
%% Creator: Inkscape inkscape 0.48.4, www.inkscape.org
%% PDF/EPS/PS + LaTeX output extension by Johan Engelen, 2010
%% Accompanies image file 'moore_t.pdf' (pdf, eps, ps)
%%
%% To include the image in your LaTeX document, write
%%   \input{<filename>.pdf_tex}
%%  instead of
%%   \includegraphics{<filename>.pdf}
%% To scale the image, write
%%   \def\svgwidth{<desired width>}
%%   \input{<filename>.pdf_tex}
%%  instead of
%%   \includegraphics[width=<desired width>]{<filename>.pdf}
%%
%% Images with a different path to the parent latex file can
%% be accessed with the `import' package (which may need to be
%% installed) using
%%   \usepackage{import}
%% in the preamble, and then including the image with
%%   \import{<path to file>}{<filename>.pdf_tex}
%% Alternatively, one can specify
%%   \graphicspath{{<path to file>/}}
%% 
%% For more information, please see info/svg-inkscape on CTAN:
%%   http://tug.ctan.org/tex-archive/info/svg-inkscape
%%
\begingroup%
  \makeatletter%
  \providecommand\color[2][]{%
    \errmessage{(Inkscape) Color is used for the text in Inkscape, but the package 'color.sty' is not loaded}%
    \renewcommand\color[2][]{}%
  }%
  \providecommand\transparent[1]{%
    \errmessage{(Inkscape) Transparency is used (non-zero) for the text in Inkscape, but the package 'transparent.sty' is not loaded}%
    \renewcommand\transparent[1]{}%
  }%
  \providecommand\rotatebox[2]{#2}%
  \ifx\svgwidth\undefined%
    \setlength{\unitlength}{109.5994175bp}%
    \ifx\svgscale\undefined%
      \relax%
    \else%
      \setlength{\unitlength}{\unitlength * \real{\svgscale}}%
    \fi%
  \else%
    \setlength{\unitlength}{\svgwidth}%
  \fi%
  \global\let\svgwidth\undefined%
  \global\let\svgscale\undefined%
  \makeatother%
  \begin{picture}(1,0.57046377)%
    \put(0,0){\includegraphics[width=\unitlength]{figures/moore_t.pdf}}%
    \put(0.22046795,0.22163112){\color[rgb]{0,0,0}\makebox(0,0)[lb]{\smash{$q_0$}}}%
    \put(0.22995357,0.09831806){\color[rgb]{0,0,0}\makebox(0,0)[lb]{\smash{$y_1$}}}%
    \put(0.22975658,0.49075308){\color[rgb]{0,0,0}\makebox(0,0)[lb]{\smash{$x_1$}}}%
    \put(0.69474893,0.22163112){\color[rgb]{0,0,0}\makebox(0,0)[lb]{\smash{$q_1$}}}%
    \put(0.70423455,0.09831806){\color[rgb]{0,0,0}\makebox(0,0)[lb]{\smash{$y_2$}}}%
    \put(0.70403756,0.49075308){\color[rgb]{0,0,0}\makebox(0,0)[lb]{\smash{$x_2$}}}%
    \put(0.45281656,0.29112034){\color[rgb]{0,0,0}\makebox(0,0)[lb]{\smash{$x_2$}}}%
    \put(0.46846061,0.01834133){\color[rgb]{0,0,0}\makebox(0,0)[lb]{\smash{$x_1$}}}%
  \end{picture}%
\endgroup%

%% file: figures/mealy_t.pdf_tex
%% Creator: Inkscape inkscape 0.48.4, www.inkscape.org
%% PDF/EPS/PS + LaTeX output extension by Johan Engelen, 2010
%% Accompanies image file 'mealy_t.pdf' (pdf, eps, ps)
%%
%% To include the image in your LaTeX document, write
%%   \input{<filename>.pdf_tex}
%%  instead of
%%   \includegraphics{<filename>.pdf}
%% To scale the image, write
%%   \def\svgwidth{<desired width>}
%%   \input{<filename>.pdf_tex}
%%  instead of
%%   \includegraphics[width=<desired width>]{<filename>.pdf}
%%
%% Images with a different path to the parent latex file can
%% be accessed with the `import' package (which may need to be
%% installed) using
%%   \usepackage{import}
%% in the preamble, and then including the image with
%%   \import{<path to file>}{<filename>.pdf_tex}
%% Alternatively, one can specify
%%   \graphicspath{{<path to file>/}}
%% 
%% For more information, please see info/svg-inkscape on CTAN:
%%   http://tug.ctan.org/tex-archive/info/svg-inkscape
%%
\begingroup%
  \makeatletter%
  \providecommand\color[2][]{%
    \errmessage{(Inkscape) Color is used for the text in Inkscape, but the package 'color.sty' is not loaded}%
    \renewcommand\color[2][]{}%
  }%
  \providecommand\transparent[1]{%
    \errmessage{(Inkscape) Transparency is used (non-zero) for the text in Inkscape, but the package 'transparent.sty' is not loaded}%
    \renewcommand\transparent[1]{}%
  }%
  \providecommand\rotatebox[2]{#2}%
  \ifx\svgwidth\undefined%
    \setlength{\unitlength}{102.6535064bp}%
    \ifx\svgscale\undefined%
      \relax%
    \else%
      \setlength{\unitlength}{\unitlength * \real{\svgscale}}%
    \fi%
  \else%
    \setlength{\unitlength}{\svgwidth}%
  \fi%
  \global\let\svgwidth\undefined%
  \global\let\svgscale\undefined%
  \makeatother%
  \begin{picture}(1,0.55789448)%
    \put(0,0){\includegraphics[width=\unitlength]{figures/mealy_t.pdf}}%
    \put(0.22534165,0.19094419){\color[rgb]{0,0,0}\makebox(0,0)[lb]{\smash{$q_0$}}}%
    \put(0.83298887,0.19094419){\color[rgb]{0,0,0}\makebox(0,0)[lb]{\smash{$q_1$}}}%
    \put(0.15484471,0.52515008){\color[rgb]{0,0,0}\makebox(0,0)[lb]{\smash{$x_1 / y_1$}}}%
    \put(0.7628873,0.52515008){\color[rgb]{0,0,0}\makebox(0,0)[lb]{\smash{$x_2 / y_1$}}}%
    \put(0.44358531,0.34522216){\color[rgb]{0,0,0}\makebox(0,0)[lb]{\smash{$x_2 / y_2$}}}%
    \put(0.44358531,0.01101628){\color[rgb]{0,0,0}\makebox(0,0)[lb]{\smash{$x_1 / y_2$}}}%
  \end{picture}%
\endgroup%

%% file: figures/dfa_t.pdf_tex
%% Creator: Inkscape inkscape 0.48.4, www.inkscape.org
%% PDF/EPS/PS + LaTeX output extension by Johan Engelen, 2010
%% Accompanies image file 'dfa_t.pdf' (pdf, eps, ps)
%%
%% To include the image in your LaTeX document, write
%%   \input{<filename>.pdf_tex}
%%  instead of
%%   \includegraphics{<filename>.pdf}
%% To scale the image, write
%%   \def\svgwidth{<desired width>}
%%   \input{<filename>.pdf_tex}
%%  instead of
%%   \includegraphics[width=<desired width>]{<filename>.pdf}
%%
%% Images with a different path to the parent latex file can
%% be accessed with the `import' package (which may need to be
%% installed) using
%%   \usepackage{import}
%% in the preamble, and then including the image with
%%   \import{<path to file>}{<filename>.pdf_tex}
%% Alternatively, one can specify
%%   \graphicspath{{<path to file>/}}
%% 
%% For more information, please see info/svg-inkscape on CTAN:
%%   http://tug.ctan.org/tex-archive/info/svg-inkscape
%%
\begingroup%
  \makeatletter%
  \providecommand\color[2][]{%
    \errmessage{(Inkscape) Color is used for the text in Inkscape, but the package 'color.sty' is not loaded}%
    \renewcommand\color[2][]{}%
  }%
  \providecommand\transparent[1]{%
    \errmessage{(Inkscape) Transparency is used (non-zero) for the text in Inkscape, but the package 'transparent.sty' is not loaded}%
    \renewcommand\transparent[1]{}%
  }%
  \providecommand\rotatebox[2]{#2}%
  \ifx\svgwidth\undefined%
    \setlength{\unitlength}{92.25732272bp}%
    \ifx\svgscale\undefined%
      \relax%
    \else%
      \setlength{\unitlength}{\unitlength * \real{\svgscale}}%
    \fi%
  \else%
    \setlength{\unitlength}{\svgwidth}%
  \fi%
  \global\let\svgwidth\undefined%
  \global\let\svgscale\undefined%
  \makeatother%
  \begin{picture}(1,0.53325378)%
    \put(0,0){\includegraphics[width=\unitlength]{figures/dfa_t.pdf}}%
    \put(0.24653088,0.12495305){\color[rgb]{0,0,0}\makebox(0,0)[lb]{\smash{$q_0$}}}%
    \put(0.80339556,0.12495305){\color[rgb]{0,0,0}\makebox(0,0)[lb]{\smash{$q_1$}}}%
    \put(0.26375355,0.49118518){\color[rgb]{0,0,0}\makebox(0,0)[lb]{\smash{$b$}}}%
    \put(0.7954286,0.49681952){\color[rgb]{0,0,0}\makebox(0,0)[lb]{\smash{$a,b$}}}%
    \put(0.54044923,0.16056086){\color[rgb]{0,0,0}\makebox(0,0)[lb]{\smash{$a$}}}%
  \end{picture}%
\endgroup%

%% file: figures/nfa_t.pdf_tex
%% Creator: Inkscape inkscape 0.48.4, www.inkscape.org
%% PDF/EPS/PS + LaTeX output extension by Johan Engelen, 2010
%% Accompanies image file 'nfa_t.pdf' (pdf, eps, ps)
%%
%% To include the image in your LaTeX document, write
%%   \input{<filename>.pdf_tex}
%%  instead of
%%   \includegraphics{<filename>.pdf}
%% To scale the image, write
%%   \def\svgwidth{<desired width>}
%%   \input{<filename>.pdf_tex}
%%  instead of
%%   \includegraphics[width=<desired width>]{<filename>.pdf}
%%
%% Images with a different path to the parent latex file can
%% be accessed with the `import' package (which may need to be
%% installed) using
%%   \usepackage{import}
%% in the preamble, and then including the image with
%%   \import{<path to file>}{<filename>.pdf_tex}
%% Alternatively, one can specify
%%   \graphicspath{{<path to file>/}}
%% 
%% For more information, please see info/svg-inkscape on CTAN:
%%   http://tug.ctan.org/tex-archive/info/svg-inkscape
%%
\begingroup%
  \makeatletter%
  \providecommand\color[2][]{%
    \errmessage{(Inkscape) Color is used for the text in Inkscape, but the package 'color.sty' is not loaded}%
    \renewcommand\color[2][]{}%
  }%
  \providecommand\transparent[1]{%
    \errmessage{(Inkscape) Transparency is used (non-zero) for the text in Inkscape, but the package 'transparent.sty' is not loaded}%
    \renewcommand\transparent[1]{}%
  }%
  \providecommand\rotatebox[2]{#2}%
  \ifx\svgwidth\undefined%
    \setlength{\unitlength}{92.25732272bp}%
    \ifx\svgscale\undefined%
      \relax%
    \else%
      \setlength{\unitlength}{\unitlength * \real{\svgscale}}%
    \fi%
  \else%
    \setlength{\unitlength}{\svgwidth}%
  \fi%
  \global\let\svgwidth\undefined%
  \global\let\svgscale\undefined%
  \makeatother%
  \begin{picture}(1,0.52198512)%
    \put(0,0){\includegraphics[width=\unitlength]{figures/nfa_t.pdf}}%
    \put(0.23777154,0.12495305){\color[rgb]{0,0,0}\makebox(0,0)[lb]{\smash{$q_0$}}}%
    \put(0.80558577,0.12495305){\color[rgb]{0,0,0}\makebox(0,0)[lb]{\smash{$q_1$}}}%
    \put(0.2615636,0.48555085){\color[rgb]{0,0,0}\makebox(0,0)[lb]{\smash{$a$}}}%
    \put(0.82812313,0.47991651){\color[rgb]{0,0,0}\makebox(0,0)[lb]{\smash{$b$}}}%
    \put(0.54044936,0.16056084){\color[rgb]{0,0,0}\makebox(0,0)[lb]{\smash{$a$}}}%
  \end{picture}%
\endgroup%

%% file: figures/pta_t.pdf_tex
%% Creator: Inkscape inkscape 0.48.4, www.inkscape.org
%% PDF/EPS/PS + LaTeX output extension by Johan Engelen, 2010
%% Accompanies image file 'pta_t.pdf' (pdf, eps, ps)
%%
%% To include the image in your LaTeX document, write
%%   \input{<filename>.pdf_tex}
%%  instead of
%%   \includegraphics{<filename>.pdf}
%% To scale the image, write
%%   \def\svgwidth{<desired width>}
%%   \input{<filename>.pdf_tex}
%%  instead of
%%   \includegraphics[width=<desired width>]{<filename>.pdf}
%%
%% Images with a different path to the parent latex file can
%% be accessed with the `import' package (which may need to be
%% installed) using
%%   \usepackage{import}
%% in the preamble, and then including the image with
%%   \import{<path to file>}{<filename>.pdf_tex}
%% Alternatively, one can specify
%%   \graphicspath{{<path to file>/}}
%% 
%% For more information, please see info/svg-inkscape on CTAN:
%%   http://tug.ctan.org/tex-archive/info/svg-inkscape
%%
\begingroup%
  \makeatletter%
  \providecommand\color[2][]{%
    \errmessage{(Inkscape) Color is used for the text in Inkscape, but the package 'color.sty' is not loaded}%
    \renewcommand\color[2][]{}%
  }%
  \providecommand\transparent[1]{%
    \errmessage{(Inkscape) Transparency is used (non-zero) for the text in Inkscape, but the package 'transparent.sty' is not loaded}%
    \renewcommand\transparent[1]{}%
  }%
  \providecommand\rotatebox[2]{#2}%
  \ifx\svgwidth\undefined%
    \setlength{\unitlength}{123.0188895bp}%
    \ifx\svgscale\undefined%
      \relax%
    \else%
      \setlength{\unitlength}{\unitlength * \real{\svgscale}}%
    \fi%
  \else%
    \setlength{\unitlength}{\svgwidth}%
  \fi%
  \global\let\svgwidth\undefined%
  \global\let\svgscale\undefined%
  \makeatother%
  \begin{picture}(1,0.72400072)%
    \put(0,0){\includegraphics[width=\unitlength]{figures/pta_t.pdf}}%
    \put(0.83798699,0.25743566){\color[rgb]{0,0,0}\makebox(0,0)[lb]{\smash{$q_{ab}$}}}%
    \put(0.83798699,0.59547102){\color[rgb]{0,0,0}\makebox(0,0)[lb]{\smash{$q_{aa}$}}}%
    \put(0.51591251,0.42809565){\color[rgb]{0,0,0}\makebox(0,0)[lb]{\smash{$q_a$}}}%
    \put(0.51591251,0.09006042){\color[rgb]{0,0,0}\makebox(0,0)[lb]{\smash{$q_b$}}}%
    \put(0.17787715,0.25907797){\color[rgb]{0,0,0}\makebox(0,0)[lb]{\smash{$q_\epsilon$}}}%
    \put(0.34455922,0.36821412){\color[rgb]{0,0,0}\makebox(0,0)[lb]{\smash{$a$}}}%
    \put(0.68259458,0.5372318){\color[rgb]{0,0,0}\makebox(0,0)[lb]{\smash{$a$}}}%
    \put(0.70583451,0.37260449){\color[rgb]{0,0,0}\makebox(0,0)[lb]{\smash{$b$}}}%
    \put(0.36779915,0.2035868){\color[rgb]{0,0,0}\makebox(0,0)[lb]{\smash{$b$}}}%
  \end{picture}%
\endgroup%

%% file: figures/ptap1_t.pdf_tex
%% Creator: Inkscape inkscape 0.48.4, www.inkscape.org
%% PDF/EPS/PS + LaTeX output extension by Johan Engelen, 2010
%% Accompanies image file 'ptm1_t.pdf' (pdf, eps, ps)
%%
%% To include the image in your LaTeX document, write
%%   \input{<filename>.pdf_tex}
%%  instead of
%%   \includegraphics{<filename>.pdf}
%% To scale the image, write
%%   \def\svgwidth{<desired width>}
%%   \input{<filename>.pdf_tex}
%%  instead of
%%   \includegraphics[width=<desired width>]{<filename>.pdf}
%%
%% Images with a different path to the parent latex file can
%% be accessed with the `import' package (which may need to be
%% installed) using
%%   \usepackage{import}
%% in the preamble, and then including the image with
%%   \import{<path to file>}{<filename>.pdf_tex}
%% Alternatively, one can specify
%%   \graphicspath{{<path to file>/}}
%% 
%% For more information, please see info/svg-inkscape on CTAN:
%%   http://tug.ctan.org/tex-archive/info/svg-inkscape
%%
\begingroup%
  \makeatletter%
  \providecommand\color[2][]{%
    \errmessage{(Inkscape) Color is used for the text in Inkscape, but the package 'color.sty' is not loaded}%
    \renewcommand\color[2][]{}%
  }%
  \providecommand\transparent[1]{%
    \errmessage{(Inkscape) Transparency is used (non-zero) for the text in Inkscape, but the package 'transparent.sty' is not loaded}%
    \renewcommand\transparent[1]{}%
  }%
  \providecommand\rotatebox[2]{#2}%
  \ifx\svgwidth\undefined%
    \setlength{\unitlength}{123.0188895bp}%
    \ifx\svgscale\undefined%
      \relax%
    \else%
      \setlength{\unitlength}{\unitlength * \real{\svgscale}}%
    \fi%
  \else%
    \setlength{\unitlength}{\svgwidth}%
  \fi%
  \global\let\svgwidth\undefined%
  \global\let\svgscale\undefined%
  \makeatother%
  \begin{picture}(1,0.55498316)%
    \put(0,0){\includegraphics[width=\unitlength]{figures/ptap1_t.pdf}}%
    \put(0.83798696,0.08841819){\color[rgb]{0,0,0}\makebox(0,0)[lb]{\smash{$q_{ab}$}}}%
    \put(0.83798696,0.42645355){\color[rgb]{0,0,0}\makebox(0,0)[lb]{\smash{$q_{aa}$}}}%
    \put(0.52248159,0.26072049){\color[rgb]{0,0,0}\makebox(0,0)[lb]{\smash{$q_a$}}}%
    \put(0.18444627,0.09170281){\color[rgb]{0,0,0}\makebox(0,0)[lb]{\smash{$q_\epsilon$}}}%
    \put(0.34455919,0.19919652){\color[rgb]{0,0,0}\makebox(0,0)[lb]{\smash{$a$}}}%
    \put(0.68259455,0.3682142){\color[rgb]{0,0,0}\makebox(0,0)[lb]{\smash{$a$}}}%
    \put(0.70583448,0.20358676){\color[rgb]{0,0,0}\makebox(0,0)[lb]{\smash{$b$}}}%
  \end{picture}%
\endgroup%

%% file: figures/ptap2_t.pdf_tex
%% Creator: Inkscape inkscape 0.48.4, www.inkscape.org
%% PDF/EPS/PS + LaTeX output extension by Johan Engelen, 2010
%% Accompanies image file 'ptm2_t.pdf' (pdf, eps, ps)
%%
%% To include the image in your LaTeX document, write
%%   \input{<filename>.pdf_tex}
%%  instead of
%%   \includegraphics{<filename>.pdf}
%% To scale the image, write
%%   \def\svgwidth{<desired width>}
%%   \input{<filename>.pdf_tex}
%%  instead of
%%   \includegraphics[width=<desired width>]{<filename>.pdf}
%%
%% Images with a different path to the parent latex file can
%% be accessed with the `import' package (which may need to be
%% installed) using
%%   \usepackage{import}
%% in the preamble, and then including the image with
%%   \import{<path to file>}{<filename>.pdf_tex}
%% Alternatively, one can specify
%%   \graphicspath{{<path to file>/}}
%% 
%% For more information, please see info/svg-inkscape on CTAN:
%%   http://tug.ctan.org/tex-archive/info/svg-inkscape
%%
\begingroup%
  \makeatletter%
  \providecommand\color[2][]{%
    \errmessage{(Inkscape) Color is used for the text in Inkscape, but the package 'color.sty' is not loaded}%
    \renewcommand\color[2][]{}%
  }%
  \providecommand\transparent[1]{%
    \errmessage{(Inkscape) Transparency is used (non-zero) for the text in Inkscape, but the package 'transparent.sty' is not loaded}%
    \renewcommand\transparent[1]{}%
  }%
  \providecommand\rotatebox[2]{#2}%
  \ifx\svgwidth\undefined%
    \setlength{\unitlength}{123.0188895bp}%
    \ifx\svgscale\undefined%
      \relax%
    \else%
      \setlength{\unitlength}{\unitlength * \real{\svgscale}}%
    \fi%
  \else%
    \setlength{\unitlength}{\svgwidth}%
  \fi%
  \global\let\svgwidth\undefined%
  \global\let\svgscale\undefined%
  \makeatother%
  \begin{picture}(1,0.55498291)%
    \put(0,0){\includegraphics[width=\unitlength]{figures/ptap2_t.pdf}}%
    \put(0.83798696,0.08841807){\color[rgb]{0,0,0}\makebox(0,0)[lb]{\smash{$q_{ab}$}}}%
    \put(0.83798696,0.42645343){\color[rgb]{0,0,0}\makebox(0,0)[lb]{\smash{$q_{aa}$}}}%
    \put(0.52248159,0.26072037){\color[rgb]{0,0,0}\makebox(0,0)[lb]{\smash{$q_a$}}}%
    \put(0.18444623,0.09170269){\color[rgb]{0,0,0}\makebox(0,0)[lb]{\smash{$q_\epsilon$}}}%
    \put(0.34455919,0.1991964){\color[rgb]{0,0,0}\makebox(0,0)[lb]{\smash{$a$}}}%
    \put(0.68259455,0.36821408){\color[rgb]{0,0,0}\makebox(0,0)[lb]{\smash{$a$}}}%
    \put(0.70583448,0.20358664){\color[rgb]{0,0,0}\makebox(0,0)[lb]{\smash{$b$}}}%
  \end{picture}%
\endgroup%

%% file: figures/white1_t.pdf_tex
%% Creator: Inkscape inkscape 0.48.4, www.inkscape.org
%% PDF/EPS/PS + LaTeX output extension by Johan Engelen, 2010
%% Accompanies image file 'white1_t.pdf' (pdf, eps, ps)
%%
%% To include the image in your LaTeX document, write
%%   \input{<filename>.pdf_tex}
%%  instead of
%%   \includegraphics{<filename>.pdf}
%% To scale the image, write
%%   \def\svgwidth{<desired width>}
%%   \input{<filename>.pdf_tex}
%%  instead of
%%   \includegraphics[width=<desired width>]{<filename>.pdf}
%%
%% Images with a different path to the parent latex file can
%% be accessed with the `import' package (which may need to be
%% installed) using
%%   \usepackage{import}
%% in the preamble, and then including the image with
%%   \import{<path to file>}{<filename>.pdf_tex}
%% Alternatively, one can specify
%%   \graphicspath{{<path to file>/}}
%% 
%% For more information, please see info/svg-inkscape on CTAN:
%%   http://tug.ctan.org/tex-archive/info/svg-inkscape
%%
\begingroup%
  \makeatletter%
  \providecommand\color[2][]{%
    \errmessage{(Inkscape) Color is used for the text in Inkscape, but the package 'color.sty' is not loaded}%
    \renewcommand\color[2][]{}%
  }%
  \providecommand\transparent[1]{%
    \errmessage{(Inkscape) Transparency is used (non-zero) for the text in Inkscape, but the package 'transparent.sty' is not loaded}%
    \renewcommand\transparent[1]{}%
  }%
  \providecommand\rotatebox[2]{#2}%
  \ifx\svgwidth\undefined%
    \setlength{\unitlength}{128.43250959bp}%
    \ifx\svgscale\undefined%
      \relax%
    \else%
      \setlength{\unitlength}{\unitlength * \real{\svgscale}}%
    \fi%
  \else%
    \setlength{\unitlength}{\svgwidth}%
  \fi%
  \global\let\svgwidth\undefined%
  \global\let\svgscale\undefined%
  \makeatother%
  \begin{picture}(1,0.89624243)%
    \put(0,0){\includegraphics[width=\unitlength]{figures/white1_t.pdf}}%
    \put(0.19405178,0.4895599){\color[rgb]{0,0,0}\makebox(0,0)[lb]{\smash{$q_0$}}}%
    \put(0.84162512,0.4895599){\color[rgb]{0,0,0}\makebox(0,0)[lb]{\smash{$q_3$}}}%
    \put(0.51783845,0.81334657){\color[rgb]{0,0,0}\makebox(0,0)[lb]{\smash{$q_1$}}}%
    \put(0.51783845,0.16577322){\color[rgb]{0,0,0}\makebox(0,0)[lb]{\smash{$q_2$}}}%
    \put(0.20761457,0.36409256){\color[rgb]{0,0,0}\makebox(0,0)[lb]{\smash{$0$}}}%
    \put(0.53140124,0.68787923){\color[rgb]{0,0,0}\makebox(0,0)[lb]{\smash{$1$}}}%
    \put(0.53140124,0.04030595){\color[rgb]{0,0,0}\makebox(0,0)[lb]{\smash{$2$}}}%
    \put(0.85518791,0.36409262){\color[rgb]{0,0,0}\makebox(0,0)[lb]{\smash{$2$}}}%
    \put(0.4406139,0.35779238){\color[rgb]{0,0,0}\makebox(0,0)[lb]{\smash{$a$}}}%
    \put(0.28681523,0.17566238){\color[rgb]{0,0,0}\makebox(0,0)[lb]{\smash{$a$}}}%
    \put(0.29864251,0.67776226){\color[rgb]{0,0,0}\makebox(0,0)[lb]{\smash{$b$}}}%
    \put(0.61506437,0.34626171){\color[rgb]{0,0,0}\makebox(0,0)[lb]{\smash{$b$}}}%
    \put(0.76960672,0.1859955){\color[rgb]{0,0,0}\makebox(0,0)[lb]{\smash{$a,b$}}}%
    \put(0.77402929,0.67848618){\color[rgb]{0,0,0}\makebox(0,0)[lb]{\smash{$a,b$}}}%
  \end{picture}%
\endgroup%

%% file: figures/white2_t.pdf_tex
%% Creator: Inkscape inkscape 0.48.4, www.inkscape.org
%% PDF/EPS/PS + LaTeX output extension by Johan Engelen, 2010
%% Accompanies image file 'white2_t.pdf' (pdf, eps, ps)
%%
%% To include the image in your LaTeX document, write
%%   \input{<filename>.pdf_tex}
%%  instead of
%%   \includegraphics{<filename>.pdf}
%% To scale the image, write
%%   \def\svgwidth{<desired width>}
%%   \input{<filename>.pdf_tex}
%%  instead of
%%   \includegraphics[width=<desired width>]{<filename>.pdf}
%%
%% Images with a different path to the parent latex file can
%% be accessed with the `import' package (which may need to be
%% installed) using
%%   \usepackage{import}
%% in the preamble, and then including the image with
%%   \import{<path to file>}{<filename>.pdf_tex}
%% Alternatively, one can specify
%%   \graphicspath{{<path to file>/}}
%% 
%% For more information, please see info/svg-inkscape on CTAN:
%%   http://tug.ctan.org/tex-archive/info/svg-inkscape
%%
\begingroup%
  \makeatletter%
  \providecommand\color[2][]{%
    \errmessage{(Inkscape) Color is used for the text in Inkscape, but the package 'color.sty' is not loaded}%
    \renewcommand\color[2][]{}%
  }%
  \providecommand\transparent[1]{%
    \errmessage{(Inkscape) Transparency is used (non-zero) for the text in Inkscape, but the package 'transparent.sty' is not loaded}%
    \renewcommand\transparent[1]{}%
  }%
  \providecommand\rotatebox[2]{#2}%
  \ifx\svgwidth\undefined%
    \setlength{\unitlength}{128.47934477bp}%
    \ifx\svgscale\undefined%
      \relax%
    \else%
      \setlength{\unitlength}{\unitlength * \real{\svgscale}}%
    \fi%
  \else%
    \setlength{\unitlength}{\svgwidth}%
  \fi%
  \global\let\svgwidth\undefined%
  \global\let\svgscale\undefined%
  \makeatother%
  \begin{picture}(1,0.89591566)%
    \put(0,0){\includegraphics[width=\unitlength]{figures/white2_t.pdf}}%
    \put(0.19710885,0.48938138){\color[rgb]{0,0,0}\makebox(0,0)[lb]{\smash{$r_0$}}}%
    \put(0.84444613,0.48938138){\color[rgb]{0,0,0}\makebox(0,0)[lb]{\smash{$r_3$}}}%
    \put(0.52077749,0.81305002){\color[rgb]{0,0,0}\makebox(0,0)[lb]{\smash{$r_1$}}}%
    \put(0.52077749,0.16571262){\color[rgb]{0,0,0}\makebox(0,0)[lb]{\smash{$r_2$}}}%
    \put(0.20790344,0.36395978){\color[rgb]{0,0,0}\makebox(0,0)[lb]{\smash{$0$}}}%
    \put(0.53157208,0.68762842){\color[rgb]{0,0,0}\makebox(0,0)[lb]{\smash{$1$}}}%
    \put(0.53157208,0.04029127){\color[rgb]{0,0,0}\makebox(0,0)[lb]{\smash{$2$}}}%
    \put(0.85524072,0.36395978){\color[rgb]{0,0,0}\makebox(0,0)[lb]{\smash{$2$}}}%
    \put(0.44081784,0.35766184){\color[rgb]{0,0,0}\makebox(0,0)[lb]{\smash{$a$}}}%
    \put(0.28707523,0.17559835){\color[rgb]{0,0,0}\makebox(0,0)[lb]{\smash{$a$}}}%
    \put(0.2988982,0.67751514){\color[rgb]{0,0,0}\makebox(0,0)[lb]{\smash{$b$}}}%
    \put(0.61520471,0.34613544){\color[rgb]{0,0,0}\makebox(0,0)[lb]{\smash{$b$}}}%
    \put(0.76969073,0.18592759){\color[rgb]{0,0,0}\makebox(0,0)[lb]{\smash{$a,b$}}}%
    \put(0.5659762,0.49253171){\color[rgb]{0,0,0}\makebox(0,0)[lb]{\smash{$a,b$}}}%
  \end{picture}%
\endgroup%

%% file: proofs/theorem1.tex
The fact that $\delta$ and $\lambda$ are total is guaranteed by the final step of the algorithm (line 49). Consistency with the training set can be proved inductively. Let $N$ denote the number of DFAs learned by the algorithm, which is equal to the number of bits required to represent an element of $O$. By definition, the Moore machine implicitly defined (by means of a synchronous product) by the $N$ prefix tree acceptors initially built by the algorithm is consistent with the training set. Assume that, before a merge operation is performed, the Moore machine implicitly defined by the (possibly incomplete) DFAs learned so far is consistent with the training set. It suffices to show that the result of the next merge operation also has this property. Suppose it does not. This means that there exists a $(\rho_I, \rho_O) \in S_{IO}$, such that $\lambda^*(q_0, \rho_I) \neq \rho_O$, which implies that in at least one of the learned DFAs, at least one state was added to the corresponding set of final states, while it should not have been (note that performing a merge operation on a DFA always yields a result accepting a superset of the language accepted prior to the merge). In other words, there is at least one learned DFA that is not consistent with its corresponding projection of the training set. However, due to the additional merge constraints that were introduced, this cannot happen, since all DFAs must be compatible with a merge in order for it to take place (line 29).

%% file: textwithlemmas.tex
It can be seen in the pseudocode of the $merge$ function (line 66) that when two states $q_u$, $q_v$ are merged in order to preserve determinism, the input word used to identify the resulting state is $min_< \{ u, v \}$, where $<$ is the total order defined in \S\ref{sec_cs}. 
When we say that $q_u$ is {\em smaller} than $q_v$ or $q_v$ {\em bigger} than $q_u$, we will mean $u = min_< \{ u, v \}$. We remark that when a blue state is merged with a red one, the latter is always smaller. This is a direct consequence of the tree-shaped nature of the initial prefix tree acceptor product, the fact that blue states are one-letter successors of red ones, and the specific order in which blue states are considered during the merging phase.

By saying that a state $q_u \in Q$ {\em corresponds} to a shortest prefix of $M$, we mean that $u \in S_P(M)$. By saying that a state $q_v \in Q$ {\em corresponds} to an element in $N_L(M)$, we mean that the state $q_v$ can be reached from $q_\epsilon$ using an element in $N_L(M)$.

$\texttt{red}$ and $\texttt{blue}$ refer to the sets of red and blue states,
as in the pseudocode of MooreMI.
Given a red state $q_u$, we will use $Merged(q_u)$ to denote the set of states that have been merged with / into $q_u$.

In the following, we assume that the training set used as input to the MooreMI algorithm is a characteristic sample for a minimal Moore machine $M$.

\begin{lemma}$ $
\label{thm:lemmaone}
\begin{itemize}
\item[(a)] Each red state corresponds to 
	an element of $S_P(M)$
	and as a consequence, to a state in $M$.
\item[(b)] Each blue state corresponds to an element of $N_L(M)$.
\item[(c)] $\forall q_u \in \texttt{red} : \forall q_v \in Merged(q_u) : \delta_m^*(q_{0\_m}, v) = \delta_m^*(q_{0\_m}, u)$.
\end{itemize}
\end{lemma}

\ifdefined \arxiv
\begin{proof}%[Lemma \ref{thm:lemmaone}] 
\input{proofs/lemma1.tex}
\end{proof}
\fi

\begin{lemma}
\label{thm:lemmatwo}
$|Q_m| \leq |Q_A|$.
\end{lemma}

\ifdefined \arxiv
\begin{proof}%[Lemma \ref{thm:lemmatwo}] 
\input{proofs/lemma2.tex}
\end{proof}
\fi

\begin{corollary}
\label{thm:corollaryone}
The previous lemmas imply the existence of a bijection $f_{iso} : Q_A \rightarrow Q_m$ such that $f_{iso}(q_u) = \delta_m^*(q_{0\_m}, u)$.
\end{corollary}

\begin{lemma}
\label{thm:lemmathree}
$\forall q_u \in Q_A : \lambda_A(q_u) = \lambda_m(f_{iso}(q_u))$.
\end{lemma}

\ifdefined \arxiv
\begin{proof}%[Lemma \ref{thm:lemmathree}] 
\input{proofs/lemma3.tex}
\end{proof}
\fi

\begin{lemma}
\label{thm:lemmafour}
$\forall q_u \in Q_A :\forall a \in I : \delta_m(f_{iso}(q_u), a) = $ $f_{iso}(\delta_A(q_u, a))$.
\end{lemma}

\ifdefined \arxiv
\begin{proof}%[Lemma \ref{thm:lemmafour}]
\input{proofs/lemma4.tex}
\end{proof}
\fi

%% file: proofs/lemma1.tex
By induction. Initially, $\texttt{red} = \{q_\epsilon\}$, $\texttt{blue} \subseteq \{q_a\ |\ a \in I\}$, and (a), (b), (c) all hold trivially. We assume they hold for the current sets of red, blue and unmarked states and will show they still hold after all possible operations performed by the algorithm:
\begin{itemize}
\item[(1)] If a state $q_v \in \texttt{blue}$ is merged into a state $q_u \in \texttt{red}$, then (a) trivially holds: The red state set remains the same, and the successors of $q_v$ are marked blue. Since they now are successors of a state corresponding to a shortest prefix (the red state $q_u$), they correspond to elements in the nucleus of $M$, so (b) holds too. Suppose now that (c) does not hold, i.e. it is $\delta_m^*(q_{0\_m}, v) \neq \delta_m^*(q_{0\_m}, u)$. Since, by the induction hypothesis, $u \in S_P(M)$  and $q_v$ corresponds to an element in $N_L(M)$, by the characteristic sample definition, there exist $(I,O)$-traces that distinguish $q_v$ and $q_u$ and prohibit their merge. But $q_v$ and $q_u$ were successfully merged, therefore $\delta_m^*(q_{0\_m}, v) = \delta_m^*(q_{0\_m}, u)$ and (c) holds.
\item[(2)] If a state $q_v \in \texttt{blue}$ is promoted to a red state, then it is distinct from all other red states. Moreover, since (i) the algorithm considers blue states in a specific order and (ii) whenever we perform a merge between two states $q_x$ and $q_y$ to preserve determinism the result is identified as $q_{min_<(x, y)}$,  $q_v$ is the smallest state distinct from the existing red states, therefore it corresponds to a shortest prefix. Its successors are now marked blue and since $q_v$ corresponds to a shortest prefix, they correspond to states in $N_L(M)$. Also, since the newly promoted red state is a shortest prefix distinct from the previous ones, it corresponds to a unique, different state in $M$. The above imply that (a) and (b) hold. Moreover, (c) trivially holds too.
\item[(3)] 
Regarding the additional state merges possibly required to maintain determinism after (1), they can occur between a red and a blue state, in which case the same as in (1) hold, between a blue state and a state that is either blue or unmarked, in which case we have what we want by the induction hypothesis, and between two unmarked states, in which case we do not need to show anything. However, we should mention here that for every pair of states being merged to preserve determinism, the two states involved necessarily represent the same state in $M$. Suppose, without loss of generality that after merging states $q_u$ and $q_v$ as in (1), states $q_{ua} = \delta^*(q_u, a)$ and $q_{va} = \delta^*(q_v, a)$ need to also be merged to preserve determinism. If $q_{ua}$ and $q_{va}$ do not represent the same state in $M$, their minimum distinguishing suffix $w = M_D(q_{ua}, q_{va})$ exists. But then, $a \cdot w$ is a distinguishing suffix for $q_u$ and $q_v$, which means that $q_u$ and $q_v$ represent different states in $M$. However, this cannot be, because, since by the induction hypothesis $u \in S_P(M)$ and $q_v$ corresponds to an element in $N_L(M)$, by the characteristic sample definition, if $q_u$ and $q_v$ were different states, $(I,O)$-traces prohibiting their merge would be present in the algorithm input. Therefore, $q_{ua}$ and $q_{va}$ represent the same state in $M$. The same argument can now be made if e.g. states $q_{uab}$ and $q_{vab}$ need to be merged to preserve determinism after $q_{ua}$ and $q_{va}$ are merged, and so on.
\end{itemize}

%% file: proofs/lemma2.tex
Suppose that $|Q_m| > |Q_A|$, i.e. there exists $q \in Q_m$ such that there is no equivalent of $q$ in $Q_A$. However, by the definition of the characteristic sample, the shortest prefix of $q$ appears in the algorithm input, and, according to Lemma \ref{thm:lemmaone}, it must eventually form a red state on its own. Therefore, there is no such state as $q$, and $|Q_m| \leq |Q_A|$ holds.

%% file: proofs/lemma3.tex
We have shown that $q_u \in Q_A$ corresponds to a unique state in $M$, specifically $\delta_m^*(q_{0\_m}, u)$. We have also shown that the algorithm is consistent with the training examples. This implies $\lambda_A(q_u) = \lambda_m(\delta_m^*(q_{0\_m}, u))$. Now, since, by definition, $f_{iso}(q_u) = \delta_m^*(q_{0\_m}, u)$, we have what we wanted.

%% file: proofs/lemma4.tex
Let $\delta_A(q_u, a) = \delta_A^*(q_\epsilon, u \cdot a) = q_v \in Q_A$. By definition, we have $f_{iso}(q_u) = \delta_m^*(q_{0\_m}, u)$ and $f_{iso}(q_v) = \delta_m^*(q_{0\_m}, v)$. In addition, $\delta_m^*(f_{iso}(q_u), a) = \delta_m(\delta_m^*(q_{0\_m}, u), a) $ $= \delta_m^*(q_{0\_m}, u \cdot a)$. But $\delta_A^*(q_\epsilon, u \cdot a) = q_v = \delta_A^*(q_\epsilon, v)$, therefore, from Lemma \ref{thm:lemmaone} (c) we have $\delta_m^*(q_{0\_m}, u \cdot a) = \delta_m^*(q_{0\_m}, v)$. Finally, $\delta_m(f_{iso}(q_u), a) = \delta_m(q_{0\_m}, u \cdot a) = \delta_m(q_{0\_m}, v) = f_{iso}(q_v) = f_{iso}(\delta_A(q_u, a))$, as we wanted.

%% file: proofs/theorem2.tex
Follows from Corollary \ref{thm:corollaryone}, Lemmas \ref{thm:lemmathree}, \ref{thm:lemmafour} and the observation that $f_{iso}(q_\epsilon) = q_{0\_m}$. The bijection $f_{iso}$ constitutes a witness isomorphism between $M$ and $M_A$.

%% file: proofs/theorem3.tex
Let $S_{IO}^{n} = \{(\rho_I^1,\rho_O^1), (\rho_I^2,\rho_O^2), \cdots , (\rho_I^{n},\rho_O^{n})\}$, for any index $n$.
Since $M$ is a minimal Moore machine, there exists at least one characteristic sample $S_{IO} = \{(r_I^1, r_O^1), (r_I^2, r_O^2), \cdots, (r_I^N, r_O^N)\}$ for it. By the hypothesis, $\forall j \in \{1, \cdots, N\} : \exists i_j$ such that
$\rho_I^{i_j} = r_I^j$. Let then $k = max_{j \in \{1, \cdots, N\}} i_j$. It is easy to see now that $S_{IO}^k = \{(\rho_I^1,\rho_O^1), (\rho_I^2,\rho_O^2), \cdots , (\rho_I^k,\rho_O^k)\}$ is a characteristic sample (as a superset of $S_{IO}$).
From the properties of characteristic samples it also follows that for any
$n\ge k$, $S_{IO}^{n}$ is also a characteristic sample (since 
in that case $S_{IO}^{n}\supseteq S_{IO}^k$).
Finally, from Theorem~\ref{thm:theoremtwo}, when MooreMI is given $S_{IO}^n$, for any $n\ge k$, as input, it will output a Moore machine isomorphic, and therefore equivalent, to $M$.

%% file: figures/ostia_t.pdf_tex
%% Creator: Inkscape inkscape 0.48.4, www.inkscape.org
%% PDF/EPS/PS + LaTeX output extension by Johan Engelen, 2010
%% Accompanies image file 'ostia_t.pdf' (pdf, eps, ps)
%%
%% To include the image in your LaTeX document, write
%%   \input{<filename>.pdf_tex}
%%  instead of
%%   \includegraphics{<filename>.pdf}
%% To scale the image, write
%%   \def\svgwidth{<desired width>}
%%   \input{<filename>.pdf_tex}
%%  instead of
%%   \includegraphics[width=<desired width>]{<filename>.pdf}
%%
%% Images with a different path to the parent latex file can
%% be accessed with the `import' package (which may need to be
%% installed) using
%%   \usepackage{import}
%% in the preamble, and then including the image with
%%   \import{<path to file>}{<filename>.pdf_tex}
%% Alternatively, one can specify
%%   \graphicspath{{<path to file>/}}
%% 
%% For more information, please see info/svg-inkscape on CTAN:
%%   http://tug.ctan.org/tex-archive/info/svg-inkscape
%%
\begingroup%
  \makeatletter%
  \providecommand\color[2][]{%
    \errmessage{(Inkscape) Color is used for the text in Inkscape, but the package 'color.sty' is not loaded}%
    \renewcommand\color[2][]{}%
  }%
  \providecommand\transparent[1]{%
    \errmessage{(Inkscape) Transparency is used (non-zero) for the text in Inkscape, but the package 'transparent.sty' is not loaded}%
    \renewcommand\transparent[1]{}%
  }%
  \providecommand\rotatebox[2]{#2}%
  \ifx\svgwidth\undefined%
    \setlength{\unitlength}{158.91218857bp}%
    \ifx\svgscale\undefined%
      \relax%
    \else%
      \setlength{\unitlength}{\unitlength * \real{\svgscale}}%
    \fi%
  \else%
    \setlength{\unitlength}{\svgwidth}%
  \fi%
  \global\let\svgwidth\undefined%
  \global\let\svgscale\undefined%
  \makeatother%
  \begin{picture}(1,0.69126658)%
    \put(0,0){\includegraphics[width=\unitlength]{figures/ostia_t.pdf}}%
    \put(0.15789229,0.20345408){\color[rgb]{0,0,0}\makebox(0,0)[lb]{\smash{$q_0$}}}%
    \put(0.68801605,0.20345408){\color[rgb]{0,0,0}\makebox(0,0)[lb]{\smash{$q_1$}}}%
    \put(0.42295417,0.46851596){\color[rgb]{0,0,0}\makebox(0,0)[lb]{\smash{$q_2$}}}%
    \put(0.16937327,0.10328529){\color[rgb]{0,0,0}\makebox(0,0)[lb]{\smash{$\epsilon$}}}%
    \put(0.43443515,0.36834717){\color[rgb]{0,0,0}\makebox(0,0)[lb]{\smash{$\epsilon$}}}%
    \put(0.69949703,0.10328534){\color[rgb]{0,0,0}\makebox(0,0)[lb]{\smash{$\epsilon$}}}%
    \put(0.35653096,0.67011447){\color[rgb]{0,0,0}\makebox(0,0)[lb]{\smash{$a / \epsilon , b / \epsilon$}}}%
    \put(0.4069689,0.00486707){\color[rgb]{0,0,0}\makebox(0,0)[lb]{\smash{$a / 02$}}}%
    \put(0.9112865,0.28050412){\color[rgb]{0,0,0}\makebox(0,0)[lb]{\smash{$a / 0$}}}%
    \put(0.63951731,0.36016065){\color[rgb]{0,0,0}\makebox(0,0)[lb]{\smash{$b / 220$}}}%
    \put(0.08645246,0.34376082){\color[rgb]{0,0,0}\makebox(0,0)[lb]{\smash{$b / 0122$}}}%
  \end{picture}%
\endgroup%